\pdfoutput=1
\PassOptionsToPackage{usenames,dvipsnames}{xcolor}
\documentclass[11pt]{article}

\usepackage[preprint]{acl}

\usepackage{times}
\usepackage{latexsym}
\usepackage{lipsum}
\usepackage{marvosym}

\usepackage[T1]{fontenc}

\usepackage[utf8]{inputenc}

\usepackage{microtype}

\usepackage{footnote}
\usepackage[many]{tcolorbox}
\BeforeBeginEnvironment{tcolorbox}{\savenotes}
\AfterEndEnvironment{tcolorbox}{\spewnotes}

\usepackage{amsmath,amssymb}%
\usepackage{amsthm}%
\usepackage{xcolor}%
\usepackage{graphicx}%
\usepackage{url}
\usepackage{enumitem}
\usepackage{setspace}
\usepackage{caption}
\usepackage{wrapfig}
\usepackage[caption=false, font=footnotesize, captionskip=0pt]{subfig}
\usepackage[breaklinks]{hyperref}
\usepackage{cleveref}

\usepackage{graphicx}
\usepackage{array}
\usepackage{multirow}
\usepackage{booktabs} 
\usepackage{pifont}

\SetLipsumParListSurrounders{\colorlet{black}{.}\color{gray}}{\color{black}}

\title{Inherent and emergent liability issues in LLM-based agentic systems:\\ a principal-agent perspective}

\author{
 \textbf{Garry A. Gabison\textsuperscript{1, 2, *}},
 \textbf{R. Patrick Xian\textsuperscript{3, 4, *}},
\\
 \textsuperscript{1}Queen Mary University of London,
 \textsuperscript{2}UC Berkeley,
 \textsuperscript{3}Certivize AI,
 \textsuperscript{4}UC San Francisco
\\
 \textsuperscript{*}Equal contribution
\\
 \normalsize{
   \Letter: \texttt{g.gabison\,@\,qmul.ac.uk}, \texttt{xrpatrick\,@\,gmail.com}
 }
}

\begin{document}
\maketitle
\begin{abstract}
Agentic systems powered by large language models (LLMs) are becoming progressively more complex and capable. Their increasing agency and expanding deployment settings attract growing attention to effective governance policies, monitoring, and control protocols. Based on the emerging landscape of the agentic market, we analyze potential liability issues arising from the delegated use of LLM agents and their extended systems through a principal-agent perspective. Our analysis complements existing risk-based studies on artificial agency and covers the spectrum of important aspects of the principal-agent relationship and their potential consequences at deployment. Furthermore, we motivate method developments for technical governance along the directions of interpretability and behavior evaluations, reward and conflict management, and the mitigation of misalignment and misconduct through principled engineering of detection and fail-safe mechanisms. By illustrating the outstanding issues in AI liability for LLM-based agentic systems, we aim to inform the system design, auditing, and tracing to enhance transparency and liability attribution.
\end{abstract}

\section{Introduction}
\label{sec:intro}







\begin{figure}[htbp]
    \centering
    \includegraphics[width=\linewidth]{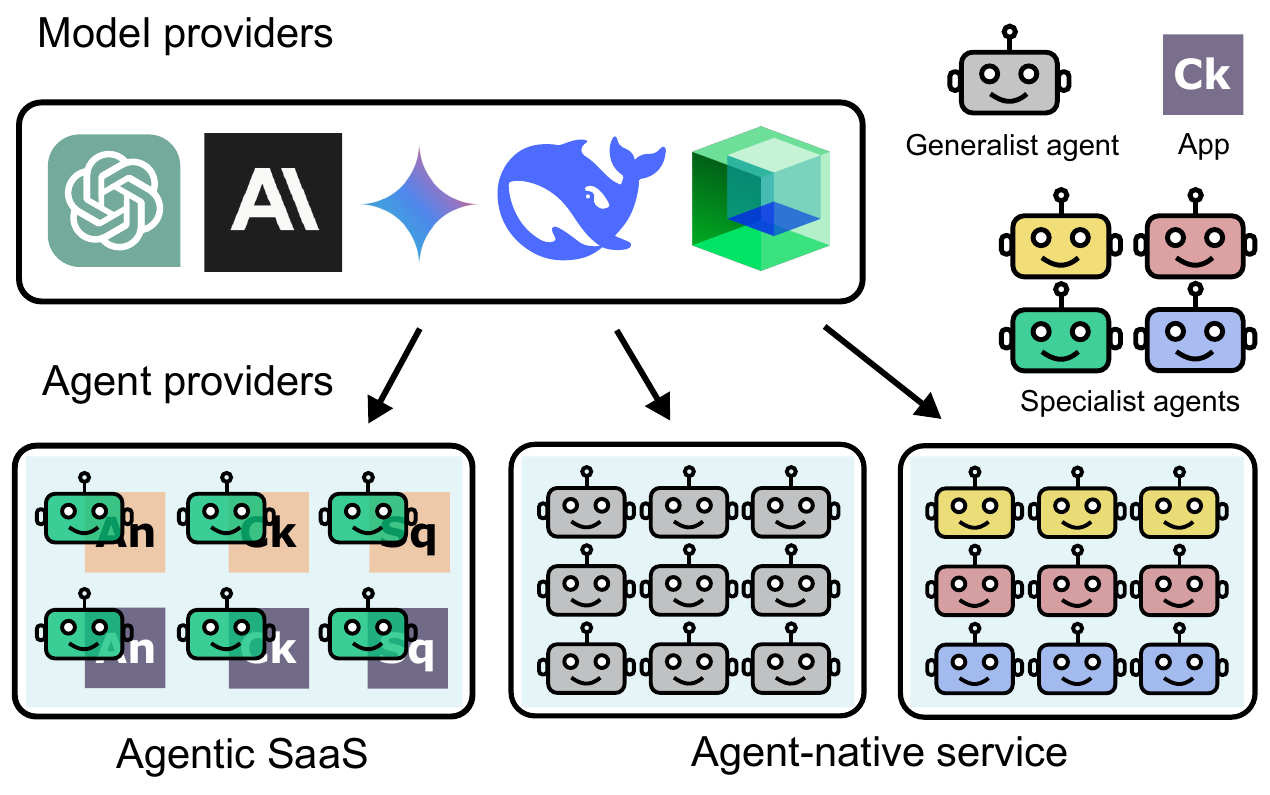}
    \caption{Landscape of the LLM agentic market. LLM agents are provided through agent-native services or agentic SaaS (agents on top of existing software apps). Agent providers are supported by model providers. Colors distinguish agents of different types.}
    \label{fig:agent_market}
    \vspace{-1em}
\end{figure}

AI agents are computer software systems capable of creating context-specific plans in non-deterministic environments \citep{chan_harms_2023,krishnan_ai_2025}. AI agents based on LLMs (aka. LLM agents, see Appendix \ref{sec:defs}) exist on a spectrum of autonomy, ranging from simple tool-calling agents to generalist agents capable of planning, sourcing, critiquing, and executing their own workflow \citep{li_llmagents_2025}. They primarily adopt an architecture with explicitly defined functioning components\footnote{Also called cognitive architecture \citep{kotseruba_cogarch_2020} at times, but the architecture alone doesn't guarantee cognition or sense of agency.} \citep{sumers_cognitive_2023}. LLM-based multiagent systems (MASs) allow agents to interact, collaborate, or compete within shared environments (Fig. \ref{fig:agent_market}a). They are designed by combining agents with specialized roles through LLM role-playing \citep{Shanahan2023,chen_persona_2024} or through integration on a software platform.
\begin{figure}[htbp]
    \centering
    \vspace{-0.5em}
    \includegraphics[width=0.9\linewidth]{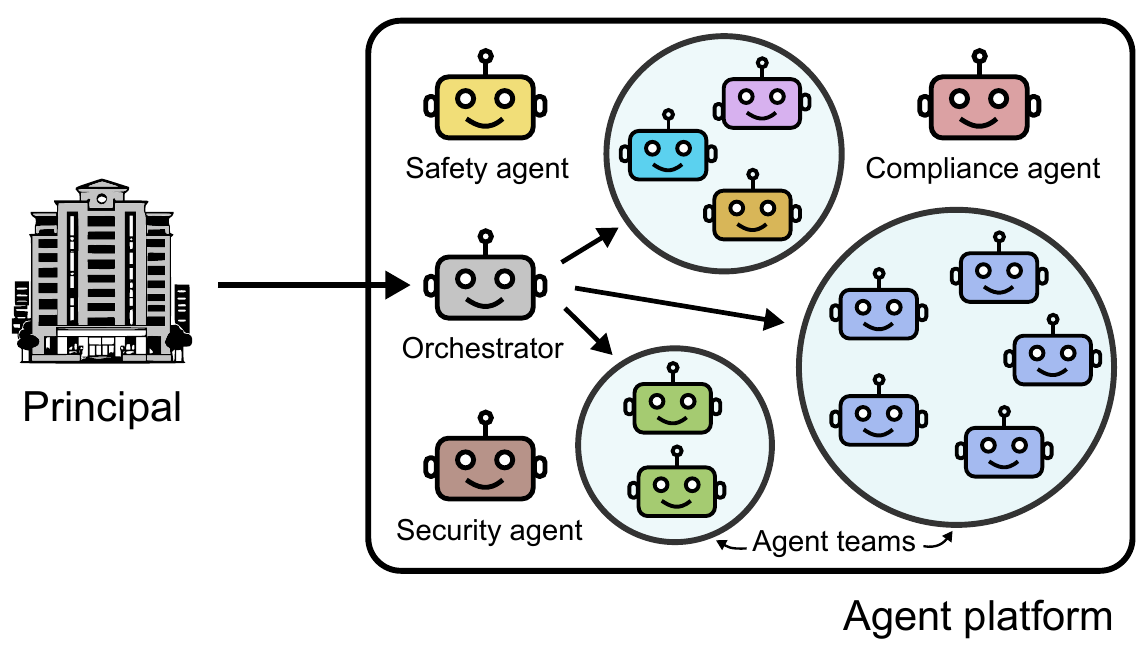}
    \caption{A plausible LLM-based MAS deployed on an agent platform, where delegation goes from the principal to an orchestrator (agent) to different functioning agent teams (circles). The platform also contains supporting agents for safety, security, and compliance. Colors distinguish between agents of different types.}
    \label{fig:agent_eg}
    \vspace{-0.5em}
\end{figure}
Each agent handles specific subtasks based on its expertise and allocated resources (tools, data, compute, etc). MASs can be tailored to a wide range of scales and domains, from few-agent systems that simulate team decision-making in medicine \citep{tang_medagents_2024,kim_mdagents_2024} and finance \citep{xiao_tradingagents_2025}, to many-agent systems that mimic the population-level socioeconomic dynamics \citep{park_generative_2024,piao_agentsociety_2025}. A plausible LLM-based MAS (Fig. \ref{fig:agent_market}a) can possess multiple teams of interacting agents coordinated by an orchestrator \citep{wang_all_2025} and behaviorally regulated by other platform agents or through a set of engineered constraints (e.g. norms) \citep{criado_open_2011,hadfield-menell_legible_2019}. The components mentioned here are further defined and explained in Appendix \ref{sec:defs}. While the flexibility of LLM-based MASs allows adaptation to various applications and affordances, it also introduces emergent risks not present in single agents \citep{hammond_multiagent_2025,pan_whymas_2025}.



Understanding the current landscape of the agentic market\footnote{\url{https://aiagentslist.com/ai-agents-map}} (Fig. \ref{fig:agent_market}b and Appendices \ref{sec:defs}-\ref{sec:providers}) is essential for analyzing liability. The present work uses the following terms to refer to its key components: Software platforms offering agent-native services include generalist agents that focuses on general autonomous use of computers, specialist agents which target labor intensive sectors and provide verticalized services for domain-specific workflow automation \citep{bousetouane_agentic_2025}, and character-infused, ``hireable'' agentic employees. Specialist agents are also offered directly by established software platforms as agentic SaaS to streamline their existing app services. Both agent-native services and agentic SaaS largely source models externally from model providers or derive their own models from open-source projects. Separately, prototypes of integration platforms and integration protocols facilitating the interaction of agents from different frameworks with third-party resources during deployment are also appearing.
  
Examples of realistic agentic systems exist across many application settings (see Table \ref{tab:agent_cases}). Because the governance of AI agents remains a nascent topic, potential liability issues in the rapidly expanding agentic market are prevalent but not sufficiently analyzed. Existing efforts are built along two streams: one focuses on establishing and refining the taxonomies of risks and harms using empirical evidence \citep{chan_harms_2023,chan_visibility_2024,he_security_2024,hammond_multiagent_2025} or differentiating the governance of agentic systems from traditional machine learning (ML) models \citep{cohen_regulating_2024,kolt_governing_2025}; another focuses on understanding the interactions between humans and AI agents to build constructive principles \citep{zheng_synergizing_2023}. Research in both streams used principal-agent theory (PAT) \citep{eisenhardt_agency_1989,laffont_theory_2002} as a starting point (see Appendix \ref{sec:pat}), but lacked a systematic examination of how existing legal frameworks for liability can effectively address various principal-agent relationships in AI systems. Moreover, AI systems are typically embedded in a sociotechnical system \citep{weidinger_sociotechnical_2023} such that the verdict on liability issues are finalized only through understanding the agent-environment interactions (see example and analysis in Appendix \ref{sec:mata}). LLM agents are still yet to be mass-deployed, so the liability issues we raise here are based on existing behavioral traits studied in technical research and hypotheticals (see Appendix \ref{app:hypothete}) extrapolated from them.

Despite the long history of PAT-based legal frameworks \citep{munday_agency_2022}, their use in AI systems is still in its early days. This work presents an initial attempt to bring principal-agent analysis to current LLM-based agentic systems. Our major contributions include: (i) extending the previous work on PAT for AI governance to accommodate the characteristic behaviors of LLM agents; (ii) discussion of the potential liability issues in light of the emerging market for LLM-based agentic systems; and (iii) outlining of the important directions of policy-driven method developments that support the technical governance of LLM agents. Without loss of generality, our analysis in the present work focuses on US jurisdictions using the Restatement of Torts and the Restatement of Law (see Appendix \ref{sec:legdef}) as the starting point to guide interpretation because tort and agency laws are not harmonized across the US. The framework put forward in this article can be applied to other jurisdictions by contextualizing the relevant agent behavior within the corresponding legal frameworks.




\section{Related works}

\paragraph{AI liability} Liability refers to the legal responsibility for one's actions. Legal scholars have treated AI as products and the harm they caused under product liability \citep{abraham2019automated,buiten_product_2024,sharkey_products_2024}.  In the US, product liability falls under: (i) design defect, (ii) manufacturing defect, and (iii) a warning defect (inadequate instructions) \citep{abraham2012forms}. Anyone harmed by a product can make a claim against any link in the supply chain.  Usually, product liability is treated under a strict liability rule, courts do not consider the care level that a manufacturer put in place to avoid an accident (see Appendix \ref{sec:legdef} for legal definitions).  Design defect can be treated under a strict liability rule or an inquiry that also resembles a negligence inquiry, courts consider the care level by considering alternative existing designs.

Service-caused harm usually triggers a negligence inquiry: was ``reasonable care'' used?  If not, the service provider is liable (primary liability) and their principal can also be liable under multiple legal theories.  Principals face two primary liabilities: ``negligent hiring'' for hiring an agent without a reasonable due diligence on its modus operandi; and ``negligent supervision'' for failing to reasonably monitor or control its agents.  Principals face secondary liability, such as vicarious liability \citep{sykes_economics_1984,diamantis_vicarious_2023}, because of their relationship (when agents acted within the boundaries of employment).  A service relationship can also fall under a contractor relationship, which does not trigger vicarious liability. Courts look at the contract and control exercised to categorize the relationship. Below, we assume the principal exercises enough control to trigger vicarious liability.

Software exhibits both ``product'' (movable goods) and ``service'' (akin to professional offering) characteristics \citep{gemignani1980product,popp_software_2011}. As AI agents become more autonomous, they move closer to services, their actions are more accountable due to increasing agenticness \citep{chopra_legal_2011,chan_harms_2023}. A negligence rule with potential vicarious liability may be more suited \citep{turner2018robot} for those relationships. In current legal use, this framework straddles product liability falling on the manufacturer and vicarious liability involving principals. Besides the service-product divide, law and economics (L\&E) has also advocated for other approaches, including risk-based liability \citep{geistfeld_comparative_2022}, fault-based liability \citep{buiten_law_2023}, explanation-based liability \citep{padovan_black_2023}, etc.

\paragraph{Principal-agent problems in AI/ML} Prior works that invoked the principal-agent framework in AI/ML primarily focused on the decision-making aspects in human-AI collaboration and AI safety. The agents here act as representatives of the principal, which differ from the agents in traditional reinforcement learning \citep{diaz_milnor-myerson_2024}. \citet{lubars_ask_2019} discussed the relation between task delegation and the principal's preference. \citet{hadfield2019incomplete} mapped AI alignment onto the principal-agent problem and discussed the alignment issues in the incomplete contracts theory. \citet{athey_allocation_2020} considered different scenarios in allocating decision authority when the human principal and the AI agent have different capabilities. \citet{critch_xrisk_2020} and \citet{hendrycks_natural_2023} considered the potential dangers of complete task delegation to agentic systems. Besides, principal-agent problems have also been considered in game-theoretic machine learning \citep{gan_generalized_2024} and in reinforcement learning settings with two interacting agents \citep{ivanov_pap_2024}.

\paragraph{Agent-oriented software systems} Agentic systems have long been proposed as a canonical approach for software design \citep{jennings_absoft_2001,zambonelli_developing_2003}. This paradigm has experienced further rise in the LLM era \citep{wang_agents_2024} because of the role-playing capability \citep{Shanahan2023,chen_persona_2024} of the these models. Their core advantages are scalability, flexibility, and the ability to perform complex tasks through task decomposition. 
LLM agents \citep{li_llmagents_2025} are configured by text instructions and they can interact with external resources to achieve enhanced capabilities than LLMs in reasoning, tool use, memory, planning, and personalization. LLM-based MASs can use verbal communication protocols to facilitate collaboration and engage in debates \citep{tran_multiagent_2025}, and the protocol topology is a key for their efficient scaling and behavioral control \citep{qian_scaling_2025}.
\begin{table*}[]
\setstretch{0.9}
    \centering
    \begin{tabular}{ccc}
    \toprule
    \textbf{Principal} & \textbf{Agentic system} & \textbf{Delegated tasks}  \\
    \midrule
    \multirow{2}{*}{Radiologist} & \multirow{2}{*}{\parbox{3.4cm}{\centering Medical imaging\\ AI agent(s)}} & \multirow{2}{*}{\parbox{8.5cm}{\setstretch{0.8}Produce a preliminary interpretation of a CT/MRI scan for a patient, suggest additional tests or treatments}} \\
    & & \\
    \midrule
    \multirow{2}{*}{\parbox{2.5cm}{\centering Frontend engineer}} & \multirow{2}{*}{\parbox{3.4cm}{\centering Website design\\ AI agent(s)}} & \multirow{2}{*}{\parbox{8.5cm}{\setstretch{0.8}Produce the code for a website from a design specification of each webpage and external media resources}} \\
    & & \\
    \midrule
    \multirow{2}{*}{Traveler} & \multirow{2}{*}{\parbox{3.4cm}{\centering Travel booking\\ AI agent(s)}} & \multirow{2}{*}{\parbox{8.5cm}{\setstretch{0.8}Plan a trip to a series of destinations including the selection of lodgings and transportation vehicles}}\\
    & & \\
    \midrule
    \multirow{2}{*}{Shopper} & \multirow{2}{*}{\parbox{3.4cm}{\centering Online shopping\\ AI agent(s)}} & \multirow{2}{*}{\parbox{8.5cm}{\setstretch{0.8}Seek and aggregate the best online deals that match items from a purchase list within a purchase budget}}\\
    & & \\
    \midrule
    \multirow{3}{*}{\parbox{2.5cm}{\centering Insurance\\ policyholder}} & \multirow{3}{*}{\parbox{3.4cm}{\centering Insurance claim\\ AI agent(s)}} & \multirow{3}{*}{\parbox{8.5cm}{\setstretch{0.8}Combine different sources of information (hospital bills, accident reports, personal emails, etc) to draft an insurance claim following a specific format restriction}}\\
    & & \\
    & & \\
    \bottomrule
    \end{tabular}
    \caption{Examples of AI agent use cases cast in the principal-agent framework. Each agentic system can be implemented using a single agent or multiple agents.}
    \label{tab:agent_cases}
    \vspace{-1em}
\end{table*}


\section{Inherent liability issues in single agents}

Contemporary approaches to the governance of AI agents \citep{kampik_governance_2022,chan_harms_2023,chan_visibility_2024,kolt_governing_2025} resort to PAT, where the principal, the human or company, delegates a task or goal to the AI agent, based on a mutual agreement. Yet LLM agents cannot satisfy all criteria of a normal agent \citep{perrier_position_2025} in PAT, creating an agency gap that can lead to an excess of unpredictable actions \citep{john_owasp_2025}. \textbf{Inherent liability issues in agentic systems arise from the dependence structure between the principal and the agent in task delegation as well as the agency gap between LLM agents and normal human agents}. We discuss these issues from the perspective of each key component of PAT (see Section \ref{sec:pat}) in the single-principal and single-agent setting.


\subsection{Artificial agency}
\label{sec:aiagency}

PAT requires a clarification of the agency relationship, which remains a hotly debated interdisciplinary topic for LLM-based systems \citep{shavit_practices_2023,dai_position_2024,barandiaran_transforming_2024,dung_understanding_2024,perrier_position_2025,mattingly_machine_2025,butlin_agency_2025,das_agency_2025}. AI researchers often take an operational view of artificial agency, such that it is possible to quantify and compare the mental state characteristics of and between AI systems through external interrogation \citep{baird_next_2021,chan_harms_2023,miehling_agentic_2025}. Representative caveats of artificial agency for LLMs and LLM agents include:
\begin{enumerate}[wide, labelindent=0pt, itemsep=-1pt, topsep=1pt]
    \item[\textbullet] \textbf{Instability}: Behavior varies with the same or paraphrased but meaning-consistent prompt on different trials \citep{loya_exploring_2023}.
    \item[\textbullet] \textbf{Inconsistency}: Behavior is sensitive to distracting contextual information or affected by sentiment and adversarially designed prompts \citep{jain_contextual_2023, maus_black_2023,zhuo_prosa_2024}.
    \item[\textbullet] \textbf{Ephemerality}: The complexity of behavioral sequence is restricted by the context window length because of the lack of effective memory mechanisms \citep{maharana_evaluating_2024}.
    \item[\textbullet] \textbf{Planning-limitedness}: Construction of executable plans hinges on accessible environmental feedback which is task-limited \citep{kambhampati_position_2024,wang_executable_2024,chen_can_2024}.
\end{enumerate}



\paragraph{Liability from flawed agency} An agency relationship requires both principals' and agents' agreement. Current LLM agents cannot yet form an authentic relationship of such because of their flawed agency \citep{barandiaran_transforming_2024,perrier_position_2025}. Voluntary relationships usually indicate that both parties benefit. However, because rationality is generally not a built-in goal in developing LLM agents \citep{macmillan-scott_irrationality_2024}, their lack of consent (from agents) means that agent providers may face liability or risks they have not considered.  That is why many have consistently argued that AI failures should be treated as product liability and because of its opacity, those failures should be adjudicated under a strict liability rule \citep{abraham2019automated,buiten_product_2024,sharkey_products_2024}.  AI providers would be left on the hook.  This approach would align with the \textit{least-cost avoidance} theory of assigning liability based on who can avoid the same accident at the lowest costs \citep{calabresi1972property} because users may not be the best place to assess whether the AI agents have any defects that could cause an accident. Even if they can, they may not be able to modify the functioning of the AI agents to avoid an accident.

The problem is that AI providers may not want to be held liable because they do not want to be exposed to unquantifiable risks as they cannot anticipate how the AI users will deploy their AI.  In such a situation, the Coase Theorem suggests that the allocation of liability and risks is best left to contracts but before the parties can do that, legislators and courts must clarify rights and liabilities \citep{coase1960problem}.  Both the AI providers and users need to use a contract to apportion possible liability or compensation mechanisms.

\subsection{Task specification and delegation}


A principal delegates tasks to their agents usually either because the principal lacks the resources or the expertise to deal with the tasks \citep{castelfranchi_towards_1998,mitchell_why_2021}. For LLM agents, the principal provides as input to the LLM a task specification that contains instructions on the nature and procedure of task execution, available resources, potential ways to overcome hurdles, the principal's preferences, etc. Such a procedure is subject to the following issues:
\begin{enumerate}[wide, labelindent=0pt, itemsep=-1pt, topsep=1pt]
    \item[\textbullet] \textbf{Task underspecification} is often present because it is impossibly costly to fully anticipate all possible scenarios to put into the specification. Moreover, the same agent may be appropriate for one task but not another; therefore, task specification -- the equivalent of job design in organizational theory \citep{oldham_job_2016} -- should balance the tradeoff between agent capability and task complexity \citep{hadfield2019incomplete}. This incompleteness leaves the door open to undesirable outcomes such as negative side effects.
    \item[\textbullet] \textbf{Risky delegation} The principal tends to forgo the delegation of very high-stakes tasks because of their severe adverse consequences and the unreliability of agent behaviors \citep{lubars_ask_2019}. The risk of delegation can be reduced by the amount of repeated feedback the principal provides to ensure alignment \citep{jiang_towards_2024}.
\end{enumerate}




\paragraph{Liability from task misdelegation} When selecting the agent (and the tasks) to delegate, principals are expected to carry out their due diligence. If they fail to take reasonable precautions when selecting agents, principals can face liability for negligently selecting \citep{camacho1993avoid}.  Negligent selection (or negligent hiring) occurs when a principal fails to exercise reasonable care when hiring an agent and the failure caused a third party to suffer harm.  Negligent selection applies beyond delegating risky tasks.  When the human principal selects the original AI agent, the original selection can trigger a negligent selection.  Once the AI agent starts selecting other AI agents to carry out subtasks, that selection is more likely to fall under a theory of negligent supervision (see Section \ref{subsec:oversight}).

Delegation of tasks also raises information concerns because the principal may not have authorized the agent to share information \citep{baird_next_2021}. Granting LLM agents access to critical information raises concerns of copyright and trademark (e.g. duplication of protected documents), trade secret (sharing information outside the system), privacy (e.g. transferring data governed by the General Data Protection Regulation or California Consumer Privacy Act), etc. That is why many tasks are not delegable \citep{mitchell_why_2021,mitchell_fully_2025}. A human agent may be able to distinguish between delegable and non-delegable tasks based on the sensitivity of the information, while an AI agent may not without clear instructions \citep{hadfield2019incomplete}.

\begin{figure*}[htbp!]
    \centering
    \includegraphics[width=0.95\linewidth]{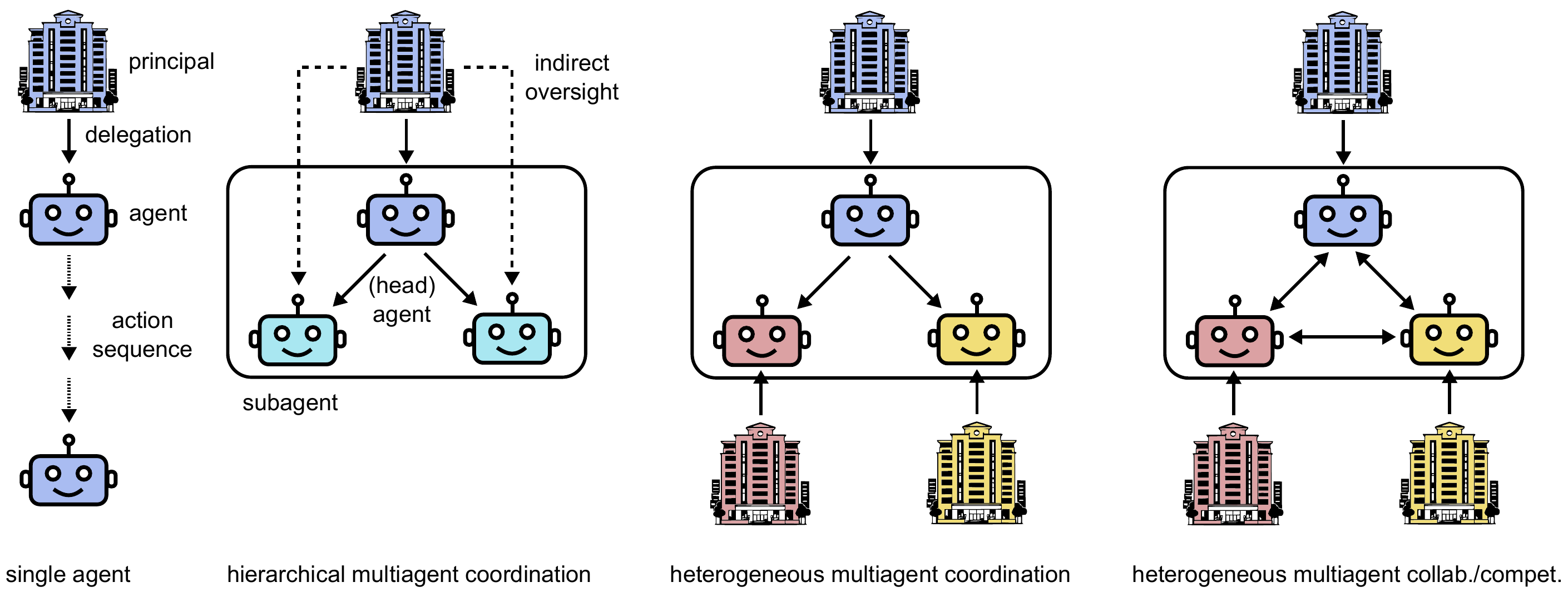}
    \caption{Examples of interaction patterns between the principals and single agents or multiagent systems (MASs). Each MAS in the coordination or collaboration (collab.)/competition (compet.) pattern is enclosed within a box. Distinct principals are colored differently.}
    \label{fig:agent_configs}
    \vspace{-1.5em}
\end{figure*}
\subsection{Principal oversight}
\label{subsec:oversight}

Human oversight is a costly endeavor for many AI applications in reality, yet it remains the gold standard in existing AI governance principles \citep{sterz_quest_2024,cihon_chilling_2024}. Principal-agent problems suffer from information asymmetries, which are usually resolved through monitoring and incentive realignment via reward or punishment. However, monitoring an AI agent requires designs that allow the principal to observe and understand what the agent is doing. Principals who prefer to realign the AI agent's incentives \citep{everitt_agent_2021} with their own would have to understand what ``motivates'' these agents or assume that they respond to human-style incentives \citep{ratliff_perspective_2019}. At the moment, quantifying misalignment remains challenging, especially for highly capable and general-purpose AI systems \citep{anwar_foundational_2024}. Moreover, the quality of principal oversight can be threatened by a spectrum of behaviors LLMs inherit from their training, such as:
\begin{enumerate}[wide, labelindent=0pt, itemsep=-1pt, topsep=1pt]
    \item[\textbullet] \textbf{Sycophancy} refers to the tendency of AI systems to provide responses that the evaluator would prefer in favor of improving the answer \citep{perez_discovering_2023,sharma_towards_2024}, exploiting the evaluator's cognitive biases (e.g. susceptibility to flattery) rather than correctly performing their duty.
    \item[\textbullet] \textbf{Manipulation} refers to the ability of LLMs to influence their principals \citep{campedelli2024want, burtell2023artificial,carroll_characterizing_2023}, towards ends that are non-welfare maximizing to their principals.
    \item[\textbullet] \textbf{Deception} refers to the tendency of AI systems to induce false beliefs \citep{park_deception_2024,scheurer_large_2024,lang_when_2024}, reinforcing the information asymmetries between the AI agent and its principal.
     \item[\textbullet] \textbf{Scheming} refers to the strategic behavior of AI systems to harbor alternative and potentially harmful motives from alignment with the principal during post-training, leading to fake alignment \citep{greenblatt_alignment_2024,balesni_towards_2024}
\end{enumerate}

These behavioral patterns also manifest in real-world scenarios \citep{blonz_costs_2023}. All of them can upend the principal-agent asymmetry such that the principal will not be able to reliably monitor the agent's behavior or provide informative feedback.

\paragraph{Liability from compromised oversight} Principals that fail to exercise oversight over agents that committed tortious acts are subject to primary liability under a negligent supervision theory \citep{cavico2016tort} act.  A principal must take reasonable care to train and supervise their agents.  If they fail to do so, they can face liability if the acts (or omissions) of their agents harm third parties or create unreasonable risks. The level of supervision depends on the context.  For example, an AI agent that is delegated the tasks of drawing up an architectural plan for a new public library, then making materials selection, and finally ordering those materials needs more supervision than an AI agent that has been put in charge of booking a holiday trip -- because the risks of harm differ.
\vspace{-0.3em}


\section{Emergent liability issues in MASs}

MASs have flexible designs (Fig. \ref{fig:agent_configs}), so agency relationships and the associated principal-agent problem can occur at different levels, similar to the functioning of a firm \citep{fama_agency_1980}. The challenges in governing single LLM agents are compounded by the participation of multiple agents in the actions, leading to additional issues with no immediate equivalent in the single-agent setting. \textbf{Emergent liability issues arise from coordination in multiagent systems and interactions between agentic systems and with supporting agents in the environment (e.g. an agent platform).} These issues exist in addition to the inherent ones for each agent. An MAS such as in Fig. \ref{fig:agent_eg} can contain an orchestrator (agent) that functions as a local manager and directs the work execution of a number of agent workers/teams with different expertise. However, imposing liability only on the orchestrator does not incentivize the improvement of the subagents\footnote{The technical distinction in designing an agentic system with fixed or adaptive roles/tasks for each agent doesn't result in different considerations of liability.}.

\subsection{Role and agency allocation}


Constructing an LLM-based MAS involves role (or task) allocation \citep{campbell_mas_2011,guo_large_2024,tran_multiagent_2025}, which implicates allocated agency, where the single agents can act on their own to accomplish the goals defined by the assigned task and resources. Because of the role-playing capability \citep{Shanahan2023,chen_persona_2024}, current LLM-based MASs generally adopt a role-centric approach, executing role allocation alongside its associated task \citep{guo_large_2024,tran_multiagent_2025}. This approach provides an interpretable division of labor and can directly mimic interactions in teams of humans. Alternatively, role allocation can be self-organized, such as in the deployment of subagents, which also need not be stationary. Role allocation is affected by the nature of LLMs' flawed agency (Section \ref{sec:aiagency}), leading to potential downstream issues:
\begin{enumerate}[wide, labelindent=0pt, itemsep=-4pt, topsep=1pt]
    \item[\textbullet] \textbf{Influenceability}: Agency of individual agents in an MAS can be enhanced or reduced through communication with other agents in a cooperative MAS \citep{he_redteaming_2025}, therefore triggering agency shift or unanticipated conduct.
    \item[\textbullet] \textbf{Distributedness}: The distribution of agency to different agents in an MAS leads to task specialization and latency which can trade off against performance and speed \citep{mieczkowski_tradeoff_2025}.
    \item[\textbullet] \textbf{Diminished control}: In a hierarchical MAS, subagents are more separated from the principal than the head agent, therefore may be harder to directly control or monitor. The principal is more prone to manipulation by the head agent.
\end{enumerate}


\paragraph{Liability from agent misallocation} These concerns make task allocation the principal's most important decision.  Firstly, the principal must decide which tasks can be delegated and to what type of agents.  A principal can face some liability for negligently selecting and supervising.  This liability exposure may deter some principal and agentic system deployment -- particularly until opacity is not resolved and human-in-the-loop is not optimized.  The other associated issue for the principals is the cost-effectiveness. As the number of allocations increases, the agency cost also increases: each new allocation must offer marginal benefits that justify its marginal costs (which also encompasses the liability exposure). From the perspective of L\&E, the liability assignment to the entity is favored if the accident can be avoided at the lowest cost \citep{calabresi1972property, carbonara2016sharing}.

Secondly, the MAS exposes its principal to the risks their (sub)agents take.  The principal must understand the risks associated with each task and with each agent.  Then, they must assess which tasks can be delegated without surpassing its risk tolerance.  However, the principal faces information asymmetries and might speculate about the expected cost-benefit of each (sub)agent. The principal also faces liability for not being careful in assigning a task to an agent, but usually not for allocating too many tasks \citep{carbonara2016sharing}. Finally, if the system is so complex or opaque, courts may decide that the harm ``speaks for itself'', inferring carelessness from the harm --- turning a negligence rule into a strict liability rule \citep{fraser2022ai, casey2019robot}.  Explainability can therefore become even more necessary.

\subsection{Operational uncertainty}

As LLM-based MASs become more complex, oversight becomes increasingly challenging.  A human overseer may only handle direct communications with the head agent, while the subsequent interactions between the head agent and the subagents are initiated autonomously among themselves. The organizational hierarchy and communication protocol can facilitate the reduction of human involvement, such as in the coordination structures in Fig. \ref{fig:agent_configs}. Unguided interactions between multiple LLM agents can create complex failure modes depending on the agent architecture and task \citep{pan_whymas_2025}, creating additional challenges in their use.
\begin{enumerate}[wide, labelindent=0pt, itemsep=-1pt, topsep=1pt]
    \item[\textbullet] \textbf{Failure cascade} refers to the scenario where the downstream agents can have increased vulnerability than the upstream agents in a MAS, which can be induced by coordination issues \citep{peigne-lefebvre_mas_2025} and communication noise induced by a confused agent \citep{barbi_preventing_2025}.
    \item[\textbullet] \textbf{Agent collusion} refers to the collaboration between agents that negatively impact others \citep{fish_algorithmic_2024,lin_strategic_2024}.
\end{enumerate}

A promising direction for minimizing operational uncertainty from misbehaving agents is to instigate corrective mechanisms and foster a norm-based governance \citep{hadfield2019incomplete,kampik_governance_2022}, where the norm is defined through spontaneous and engineered social interactions between agents \citep{trivedi_altared_2024}.

\paragraph{Liability from operational uncertainty} When multiple LLM agents interact autonomously, the attribution of responsibility becomes blurred because decisions emerge from collective behaviors rather than individual actions. A software provider usually bears the liability for the harm caused by their LLM agents:\footnote{This problem is already present in the autonomous vehicle context and has pushed governments like the UK to change their laws to shield AI providers from liability see UK Automated and Electric Vehicles Act 2018 and UK Automated Vehicles Act 2024 \citep{soder_levels_2024}.} if an AI agent is considered a product that caused harm to a third party, the victim could sue the product manufacturer (i.e., software provider here) under product liability theory, the court would usually apply a strict liability rule and the manufacturer will be held liable if the product is considered to have caused the harm \citep{turner2018robot, Furr_2024, barfield2018liability}; if an AI agent is considered a service that caused the harm to a third party, the victim could sue the service provider (i.e., software provider), the court would usually apply a negligence rule (although strict liability may also apply in risky activity e.g., putting an AI in charge of handling dynamites \citep{reid_liability_1999}), and if the service was provided negligently, the victim will recover \citep{ramakrishnan2024us, barfield2018liability}.  The question of whether a piece of software is a product or service remains a question of fact.\footnote{See e.g., \textit{Lemmon v. Snap, Inc}., 995 F. 3d 1085 (9th Cir. 2021); \textit{Holbrook v. Prodomax Automation LTD}., Case No. 1:17-cv-219 (W.D. Mich. 2021).}  Attempting to assert that an AI system is a distinct entity has failed in the past and will likely fail in the future \citep{lior2024holding}. This doesn't mean that no one else could be held liable or that the software providers cannot seek contribution under a contract clause, but that they will be usually held liable if their AI agent is deemed to have caused harm to someone to whom they owned a duty of care.

For a heterogeneous MAS, courts may apply separate liabilities or distribute liability among providers according to the harm contribution.  Courts do not favor this approach because it is complex to estimate, so the parties would benefit from contractual clarification between all the (sub)agents (e.g. contract liability based on value).  Instead, when causes cannot be disentangled and assigned, courts revert to joint and several liability such that each party is liable for the full harm and can be sued individually, which expands liability to the (sub)agents -- a tempting approach to avoid complex litigations and battle of experts when MASs are involved \citep{custers2025liability}. Courts may decide that multiagent behaviors, such as cascading failures or agent collusion tendencies, lead to third-party harm too often. So, courts may elevate the use of MASs to a ``risky activity'' and use a strict liability rule \citep{vcerka2015liability}.

\subsection{Platform integration}

Emerging LLM-based MASs feature provider-dependent agent frameworks, which will likely follow somewhat different safety protocols (Fig. \ref{fig:agent_configs}) between agent providers. At the moment, efforts to integrate different agent frameworks analogous to traditional software integration \citep{bass_software_2021} are still lagging but are expected to ramp up due to market growth. The motivation to integrate LLM agents is the enhancement of system capability by unifying disparate provider frameworks, which may include privileged access to customized agentic components (e.g. unique databases, fast memory, etc). Overall, integration serves the needs of the user (i.e. principal) by balancing the advantages of different agent frameworks as well as providing an extra layer of control and oversight through the inclusion of what we call \textit{platform agents} (see Fig. \ref{fig:agent_eg}). At the moment, the potential benefits of integration platforms for LLM agents include:
\begin{enumerate}[wide, labelindent=0pt, itemsep=-1pt, topsep=1pt]
    \item[\textbullet] \textbf{Platform oversight} refers to measures on a platform to provide users with enhanced multiagent security through a security-guard agent \citep{xiang_guardagent_2024}, collusion mitigation mechanism \citep{foxabbott_defining_2023}, detection and suppression of copyright infringement \citep{liu_copyjudge_2025} or privacy leakage. These oversight mechanisms are provided through the integration platform as platform agents that interacts with an existing agentic system.
    \item[\textbullet] \textbf{Platform teaming} refers to the formation of agent teams on an integration platform through user-defined or ad hoc protocols \citep{mirsky_survey_2022,wang_nagent_2024} that enhance cooperation among homogeneous or heterogeneous agents.
\end{enumerate}


\paragraph{Liability from mismanaged platforms} An integrated multiagent platform could carry some liability depending on the level of control it exercises for the digital entities \citep{gabison_platform_2020,lefouili_economics_2022}. For example, control might encompass the behavioral monitoring of the individual agents operating on the platform. Because the platform intrinsically involves multiple principals and multiple agents, the principals may be liable for engaging in collusive behavior but, in rare occasions, platforms have faced liability for incentivizing others (e.g. copyright infringement\footnote{\textit{MGM Studios, Inc. v. Grokster, Ltd.}, 545 U.S. 913 (2005)}, intentional interference\footnote{\textit{hiQ Labs, Inc. v. LinkedIn Corp.}, 938 F. 3d 985 (9th Cir. 2019)}).





\section{Policy-driven technical development}

The attribution of liability benefits from in-depth failure analysis of system behavior and transparency mechanisms that supports the tracing of agent actions. These in turn motivates concomitant developments in the approaches to manage agent behavior. We discuss a few directions to this end: 


\paragraph{Interpretability and behavior evaluations} Tracing an AI agent's actions \citep{lu_agentlens_2024} can be the basis for establishing whether it took reasonable care \citep{price_potential_2019,choi_software_2020} and therefore the evidence for liability claims. Prior works on the interpretability and faithfulness in LLM reasoning \cite{lyu_faithful_2023,wei_jie_how_2024} and dialogue generation \citep{tuan_local_2021} may eventually also be used to assist in the analysis of agent behavior. This analysis can be used to investigate whether the AI intended some actions -- a necessary element in intentional torts.  More generally, these kinds of evaluations may help understand unreliable behavioral patterns and help diagnose and refine the design bottlenecks in LLM agents. 


To better ground notions of reasonable care for AI agents, method developments should prioritize decomposing complex multiagent interactions into interpretable causal mechanisms, leveraging causal abstraction frameworks \cite{geiger_causal_2024} to create faithful, human-intelligible representations of agent interactions that preserve essential causal relationships while abstracting away unnecessary details.
Additionally, formal verification approaches \citep{zhang_fusion_2024} may be able to detect and prevent potential failure modes in agent interactions to improve decision-making.




\paragraph{Reward and conflict management} Some application settings of LLM-based MASs aim to mimic the functioning of human teams and organizations \citep{xie_can_2024}. Agentic systems can learn from existing organizational theory \citep{mitnick_theory_1992,vardi_misbehavior_2016} to improve the design and architecture given the flawed agency of its components. The most relevant aspects include systems for managing reward and conflict between agents. It has been shown that ``verbal tipping'' \citep{salinas_butterfly_2024} can provide concrete incentives in instruction to improve LLM performance. Moreover, managing knowledge conflict in LLMs \citep{xu_knowledge_2024} has been the most related area. To manage generic conflicts between LLM agents, a credit system \citep{thomas_credit_2017} of refusal and sanction based on agent IDs may be beneficial \citep{chan_infrastructure_2025}. This could include adaptive trust scoring from evaluation of domain-specific expertise, and an arbitration protocol to adjudicate conflicts such that the arbiter holds the right to refuse action from a frequently misbehaving agent. While the scoring system would need to be supported across the relevant AI agent integration platforms (Fig. \ref{fig:agent_market}a), it is essential to balance individual agent utility and cooperation.





\paragraph{Misalignment and misconduct avoidance} As discussed in \Cref{subsec:oversight}, LLMs and the agents based on them may act in misaligned ways, with behaviors ranging from sycophancy and deception to scheming. While AI model providers may be able to partially reduce these problems through effective detection and behavior steering, which has been demonstrated on LLMs \citep{rimsky_steering_2024,gdill_detecting_2025,williams_targeted_2025}. In LLM-based agentic systems, the equivalent tasks could be carried out using separate LLM agents through observing and analyzing other agents' behavior, such as using the theory-of-mind capability \citep{street_llm_2024}. The MAS can include agents with a specially finetuned base model \citep{binz_turning_2023} as a warden (agent) for deception mitigation. Similar approaches can also be used to suppress other agent misconduct such as the generation of harmful or copyright-protected content using the warden to filter through key tokens and phrases that can induce such behavior. Recently, \citet{hua_trustagent_2024} implemented a safety inspection step with a specialized LLM agent to improve operational safety. Such adaptive approach carried out by a separate agent can compensate for the limitations in existing model-level approaches such as machine unlearning \citep{bourtoule2021machine}, which is suffering from limited effectiveness and tradeoffs with other model capabilities \citep{cooper_machine_2024,liu_rethinking_2025}.


\section{Conclusion}

We examined liability issues arising from LLM-based agentic systems by analyzing and situating distinct aspects of the agentic AI ecosystem according to the principal-agent theory. Although a varieties of issues of LLM agents are yet to present themselves in concrete real-world examples, the growing evidence and demonstrations in simulated scenarios can inform their potential impact at the societal scale, which we built on in our prospective study. Our work shows that besides increasing agency, disruption in other aspects of the principal-agent relationship can also lead to liability incidents. Ultimately, the materialization of liability issues in reality will be dominated by the incidents that occur in the more frequent use cases of agentic systems in each industry sector. Our analysis enriches existing contextualization of AI risk \citep{chan_harms_2023,hammond_multiagent_2025} and demonstrates the explanatory power of the behavior-centric approach to translating frontier AI research into tangible knowledge to inform legal analysis and policy.






\section*{Limitations}

The present work focused on legal liabilities but did not address the potential moral responsibility issues present in LLM-based agentic systems. Although liability depends on the legal system and therefore varies across jurisdictions, the principle of vicarious liability for the actions of an agent remains consistent in most jurisdictions. The discussions were set for relatively small-scale agentic systems (e.g. up to tens of agents), where each agent has its detailed roles and tasks. Because of the execution cost and reliability issues, we don't think larger-scale MASs (e.g. having a hundred or more agents) will become practical solutions and deployed widely or in long-running instances any time soon. Their use tends to be more relevant for academic research and they have different operating conditions, including more common role/task underspecification and stronger emergent characteristics. Answering questions about failure modes and attributing liability would require more understanding of the deployment history. Another limitation is that our discussion centered around LLM agents, but the same liability issues described here are likely applicable to multimodal AI agents.

\section*{Ethics statements}

The design and deployment of LLM-based agentic systems raise significant ethical and legal considerations regarding liability attribution. This work acknowledges the complex interplay between provider responsibility, user actions, and the emergent behaviors of semi-autonomous systems powered by LLMs. We recognize that traditional liability frameworks may inadequately address scenarios where AI agents make consequential decisions with incomplete human oversight. Our analysis aims to contribute to the discussion on appropriate liability for all sides related to LLM-based agentic systems that balances current evidence from system behavior and the available legal framework. We have considered liability in different deployment environments with the aim of informing policy discussions and technical developments to make liability more traceable in AI systems that are complex and extensible, but prone to misbehavior and failure and lack in self-control.

\section*{Acknowledgments}
We thank the organizers of the AI Governance at the Crossroads symposium held at Berkeley, California in February, 2025. We thank Micah Carroll at the University of California, Berkeley for information about LLM behaviors and AI safety and helpful comments on the manuscript.




\appendix

\section{Agentic system definitions}
\label{sec:defs}

We clarify here the definitions and scopes for some of the key terms used throughout the text. They are not meant for a complete characterization of these terms but are primarily aimed at illustrating their relationships in the context of this work.

\begin{enumerate}[wide, labelindent=0pt, itemsep=0pt, topsep=1pt]
    \item[\textbullet] An \textbf{AI agent} is a generic term referring to a software agent powered by any form of artificial intelligence \citep{krishnan_ai_2025}. An equivalent definition is provided in Section \ref{sec:intro}.
    \item[\textbullet] An \textbf{LLM agent} refers specifically to an AI agent powered by at least one LLM as the central component that executes planning, initiates and coordinates the agent's actions, etc. An \textbf{LLM-based multiagent system} constitutes of multiple LLM agents that can be derived from the same or different LLMs.
    \item[\textbullet] An LLM-based MAS consists of some components that facilitate its operation. This includes \textbf{agent teams} that consist of several interacting agents, which mimics human communication in teamwork and group decision-making. An \textbf{orchestrator} is an agent that distributes tasks to other specialized agents and facilitates their actions \citep{bhatt_when_2025}. On an agent or integration platform, the platform maintainers can order specialized agents, or \textbf{platform agents}, to oversee and inspect \citep{hua_trustagent_2024} the communications between agents. They are generally aimed at improving the safety \citep{xiang_guardagent_2024}, security, and policy compliance of the actions carried out on the platform.
    \item[\textbullet] An \textbf{AI system} is an umbrella term defined in the EU AI Act\footnote{\url{https://artificialintelligenceact.eu/article/3/}}. When deployed, an AI system can operate with various levels of autonomy and can create recommendations and content, make predictions and decisions that influence the environment. The system in the term indicates that the AI is not acting in isolation, but is assisted by the surrounding infrastructure, such as cloud computing, databases, and user interfaces, which are integral to the AI system and essential for its use. An AI system can be agentic or non-agentic.
    \item[\textbullet] An \textbf{agentic system}, also known as an \textbf{agenic AI system} \cite{shavit_practices_2023}, is a type of AI system that contains a level of agency to carry out actions on its own in pursuit of a goal. An agentic system can include a single agent or multiple agents acting in coordination, competition, cooperation (or collaboration). An AI agent is a key component of an agentic system.
    \item[\textbullet] Analogous to the previous entry, \textbf{agentic market} is the segment of AI market represented by the providers and buyers of agentic AI systems.
    \item[\textbullet] An \textbf{agent platform} provides resources and toolkits to construct, configure as well as deploy AI agents (Fig. \ref{fig:agent_market}a). AI agents from distinct providers can operate on a (software) \textbf{integration platform}, where they can interact with each other, with third-party data sources, proprietary APIs, etc. AI agents on any deployment platform may be subject to compliance governance and can receive protection against cyberthreats or malfunction from the software infrastructure there.
\end{enumerate}


\section{Types and examples of existing LLM agent providers}
\label{sec:providers}

We present here preliminary examples for elements of the agentic market that are currently available.

\subsection{Agentic software as a service (SaaS)}
\begin{enumerate}[wide, labelindent=0pt, itemsep=-1pt, topsep=1pt]
    \item[\textbullet] Salesforce: \href{https://www.salesforce.com/agentforce/}{www.salesforce.com/agentforce/} 
    \item[\textbullet] Adobe: \href{https://business.adobe.com/products/experience-platform/agent-orchestrator.html}{business.adobe.com/products/experience-platform/agent-orchestrator.html}
    \item[\textbullet] SAP: \href{https://www.sap.com/products/artificial-intelligence/ai-agents.html}{www.sap.com/products/artificial-intelligence/ai-agents.html}
    \item[\textbullet] Oracle: \href{https://www.oracle.com/artificial-intelligence/generative-ai/agents/}{www.oracle.com/artificial-intelligence/generative-ai/agents/}
    \item[\textbullet] Cisco Webex: \href{https://www.webex.ai/ai-agent.html}{www.webex.ai/ai-agent.html}
\end{enumerate}
\subsection{Agent-native service}
\paragraph{Providers of generalist agents}
\begin{enumerate}[wide, labelindent=0pt, itemsep=-1pt, topsep=1pt]
    \item[\textbullet] OpenAI Operator: \href{https://operator.chatgpt.com/}{operator.chatgpt.com}
    \item[\textbullet] Google DeepMind Project Astra: \href{https://deepmind.google/technologies/project-astra/}{deepmind.google/technologies/project-astra}
    \item[\textbullet] Manus: \href{https://manus.im/}{manus.im}
    \item[\textbullet] Simular: \href{https://www.simular.ai/}{simular.ai}
\end{enumerate}

\paragraph{Providers of specialist agents}
\begin{enumerate}[wide, labelindent=0pt, itemsep=-1pt, topsep=1pt]
    \item[\textbullet] Sesame (\href{https://www.sesame.com/}{www.sesame.com}) offers voice AI agents for different domains.
    \item[\textbullet] Contextual (\href{https://contextual.ai/}{contextual.ai}) offers specialized AI agents with advanced retrieval-augmented features. 
    \item[\textbullet] Devin (\href{https://devin.ai/}{devin.ai}) offers AI agents for coding.
    \item[\textbullet] Sierra (\href{https://sierra.ai/}{sierra.ai}) offers AI agents tailored for different types of customer services.
    \item[\textbullet] Health Force (\href{https://www.healthforce.ai/}{www.healthforce.ai}) offers human resources AI agents to handle digital text processing tasks in healthcare systems.
    \item[\textbullet] Zenity (\href{https://www.zenity.io/}{www.zenity.io/}) offers security-focused AI agents.
\end{enumerate}

\paragraph{Providers of character-infused agents}
\begin{enumerate}[wide, labelindent=0pt, itemsep=-1pt, topsep=1pt]
    \item[\textbullet] Artisan: \href{https://www.artisan.co/}{www.artisan.co}
    \item[\textbullet] Sintra: \href{https://sintra.ai/}{sintra.ai}
\end{enumerate}

\subsection{Elements of agent integration platforms}

\begin{enumerate}[wide, labelindent=0pt, itemsep=-1pt, topsep=1pt]
    \item[\textbullet] Microsoft Copilot Studio: \href{https://www.microsoft.com/microsoft-copilot/microsoft-copilot-studio}{www.microsoft.com/microsoft-copilot/microsoft-copilot-studio}
    \item[\textbullet] Anthropic Model Context Protocol: \href{https://modelcontextprotocol.io/}{modelcontextprotocol.io}
    \item[\textbullet] Google Agent2Agent protocol: \href{https://google-a2a.github.io/A2A/}{google-a2a.github.io/A2A/}
    \item[\textbullet] IBM Bee AI: \href{https://beeai.dev/}{beeai.dev}
\end{enumerate}

\section{Principal agency and liability}
\label{sec:pat}

PAT examines the relationship where one party, the principal, delegates authority to another, the agent, creating three fundamental challenges \citep{eisenhardt_agency_1989,laffont_theory_2002}:
\begin{enumerate}[wide, labelindent=0pt, itemsep=-1pt, topsep=1pt]
    \item[\textbullet] \textbf{Adverse selection} (aka. hidden information problem) occurs when agents possess more information than principals about their abilities or efforts.  Adverse selection is a type of information asymmetry also known as the hidden information problem \citep{akerlof1970market}.
    \item[\textbullet] \textbf{Moral hazard} (aka. hidden action problem) occurs when agents take greater risks than principals would prefer because they do not bear the full consequences \citep{arrow1963uncertainty}. Moral hazard is also a type of information asymmetry.
    \item[\textbullet] \textbf{Misaligned interest} (aka. conflict of interest) can manifest between principals and agents in four ways \citep{jensen_theory_1976}: principal-agent collusion against third parties, principal-third party collusion against agents, agent-third party collusion against principals, or agents simply pursuing self-interest independently.  Misaligned interests lead to agency cost, which is associated with information sharing, monitoring of the agent, etc \citep{fama_agency_1983}. 
\end{enumerate}

Each of these problems has legal and economic solutions.  Legal solutions include fiduciary duties (duty of care, duty of loyalty, etc.). Those duties provide principals with a legal recourse against agents who breached those duties.  However, legal enforcement is probabilistic and slow \citep{levmore1990probabilistic}, so principals often prefer to use economic mechanisms, which depend on the problem.  Principals can address some hidden information problems by creating separating equilibria that force agents to reveal information about themselves, that is, ``signals''; signals include credentials, warranties, or performance histories that distinguish high-quality from low-quality agents \citep{spence1973job}.  Principals mitigate a hidden action problem through monitoring (i.e. direct observation of agent behavior) and then linking compensation to observable effort, or bonding arrangements in which the principals tie remuneration to outcomes (bonuses), thereby aligning financial incentives \citep{holmstrom1979moral}.  A principal can address conflicts of interest by realigning incentives through carefully designed contracts (deferred compensation), creating organizational structures that promote incentive realignment (profit sharing), or leveraging reputation mechanisms. A principal has incentives to monitor their agents.

The primary liability faced by agents can motivate them to take reasonable care when carrying out a task.  The question is whether AI agents or their providers face any liability in such a system.  Primary and secondary liability exposures incentivize principals to ensure that their agents complete the tasks with reasonable care. Overall, the ultimate principal is always a human and may be held liable for the actions of the AI agents.

\section{Legal definitions and hypotheticals}
\label{sec:legdef}

\subsection{Tort law-related terms}
In simple terms, \textbf{tort law} is the set of laws that attempt to redress civil wrongdoings which ensue from the actions (or omissions) of an individual that cause harm to a third party.  Tort law depends on the jurisdiction (i.e., jurisdiction-specific). In the United States, tort law remains the domain of the states. In the European Union, each member state sets its own tort law. A recent effort to harmonize tort law has failed to progress in the European Parliament because of the inability of member states to come to an agreement \citep{libert_ai_2025}. The definitions below are based on the various Restatements of Tort Law.  These Restatements are treatises written by American scholars and published by the American Law Institute.  They summarize tort concepts and guide courts and lawyers in the area.  The various Restatements \citep{rest_tort_2,rest_tort_3} are not binding laws but provide a good indication of what courts consider in the United States.

\begin{enumerate}[wide, labelindent=0pt, itemsep=0pt, topsep=1pt]
    \item[\textbullet] Tort law does not require the actor and the third party to have any relationship or a legally recognized privity (which, by contrast, is required for contract claims).  Every time an individual (or entity) acts, they create the chance of harming someone.  The \textbf{intent} of the actor is generally not relevant in tort law to decide whether someone is responsible for most torts. However, intent can affect whether to impose punitive damages, which are damages that go beyond compensating for the harm caused.\footnote{\textit{Restatement (Third) of Torts: Liability for Physical and Emotional Harm § 1}} 
    \item[\textbullet] In tort law, \textbf{liability} refers to a court finding that an individual or entity is responsible for the harm inflicted on a third party.  To prove that an individual (or entity) is liable, the victim must provide different evidence depending on the rule.  The most common rules are strict liability, the negligence rule, and comparative negligence.
    \item[\textbullet] The \textbf{strict liability rule} is a type of liability rule where the victim-cum-plaintiff must demonstrate to a judge (or jury) that the action-taker-cum-defendant took actions and that the actions caused the defendant harm that can be redressed usually either through a monetary compensation called damages or a court order called injunction to stop their harmful actions.  The strict liability rule ignores the level of care the defendant took in avoiding the accident.  Courts usually use a strict liability for abnormally dangerous activities\footnote{\textit{Restatement (Third) of Torts: Liability for Physical and Emotional Harm Ch. 4}} or for many product liability questions \footnote{\textit{Restatement (Third) of Torts: Product Liability § 1 1998.}} 
    \item[\textbullet] The \textbf{negligence rule} is a type of liability rule where the victim-cum-plaintiff must demonstrate to a judge (or jury) that the action-taker-cum-defendant had a duty toward the victim, failed to live up to that duty (i.e., did not take reasonable care), and that their actions caused the victim redressable harm.  The negligence rule attempts to ensure that potential tortfeasors take reasonable efforts to lower the risk imposed on others. For example, ``an actor ordinarily has a duty to exercise reasonable care when the actor's conduct creates a risk of physical harm.''\footnote{\textit{Restatement (Third) of Torts: Liability for Physical and Emotional Harm § 7(a)}}
    \item[\textbullet] When \textbf{comparative negligence rules} are used, courts weigh the responsibility of both the plaintiff and the defendant in causing the accident and assign to each litigant a percentage of responsibility for the accident.  Most courts reduce the recoverable damages by the percentage of fault attributed to the plaintiff (an approach known as ``pure comparative negligence'') or prevent all recovery if the plaintiff is more at fault than the defendant (``modified or partial comparative negligence'').
    \item[\textbullet] \textbf{Product liability} refers to the liability traced from the harm caused by a product back to its manufacturer.  Product liability occurs when a product is defective and “A product is defective when, at the time of sale or distribution, it contains a manufacturing defect, is defective in design or is defective because of inadequate instructions or warnings.''\footnote{\textit{Restatement (Third) of Torts: Product Liability § 2}} The manufacturing defect occurred because the normal manufacturing process was not followed and this deviation led to the product presenting harm, while a product manufactured using the normal process does not.  The deviation affects only one product (or a batch of products) that has been put into the stream of commerce.
    
    By contrast, a design defect and a warning defect affect all the products that have been put into the stream of commerce.  A design defect occurred when the manufacturer created a product with a feature that presents more risk than benefit.  A common test to assess whether a design defect occurred is the risk-utility test, which requires the plaintiff to show that the manufacturer could have used a reasonable alternative design, that the alternative design is safer than the marketed product, and that the alternative design was reasonably economically feasible and practical.  Finally, a warning defect or information defect occurs because certain hidden risks cannot be designed away, so the manufacturer must provide information to warn the consumers, but the information was not adequate because it was not visible, comprehensive, only used an icon, and no risk mitigation information was provided.
\end{enumerate}

\subsection{Agency law-related terms}
In simple terms, agency law governs the duties and obligations that arise in an agency relationship.  Once again, the definitions below rely on the various Restatements of Law, including the Restatement of Agency \citep{rest_agency_3} because agency law is specific to the jurisdiction.

\begin{enumerate}[wide, labelindent=0pt, itemsep=0pt, topsep=1pt]
    \item[\textbullet] An \textbf{agency relationship} is established when the ``principal'' (an individual or entity) agrees that the ``agent'' acts (another individual or entity) on their behalf, the principal can control the agent's activities, and the agent agrees to the relationship.\footnote{\textit{Restatement (Third) of Agency (2006) § 1.01}}
    \item[\textbullet] In many cases, \textbf{control} is often the key factor that courts investigate to identify whether two individuals entered into an agency relationship.  Courts do not need to find that the principal exercised control and only need to find that the principal had the ability to do so. Even if the principal exercises control and provides detailed instructions, the agent may not act in the way the principal expected because agents must interpret those instructions.
\end{enumerate}

The existence of an agency relationship triggers various types of liabilities. The ones discussed above include:
\begin{enumerate}[wide, labelindent=0pt, itemsep=-1pt, topsep=1pt]
    \item[\textbullet] \textbf{Negligent hiring} is a form of primary liability --- because it is based on the actions of the principal --- where the principal is held responsible for selecting an agent that was likely to cause harm to a third party and put them in a position that creates the chance to cause that harm.\footnote{\textit{Restatement (Third) of Agency § 7.03 (2006)}}  For example, a store manager could be held liable for negligent hiring for hiring someone known for excessive drinking as a delivery person who ends up assaulting customers in their home.\footnote{ \textit{Fleming v. Bronfin}, 80 A.2d 915 (D.C. 1951)}. The principal has a duty to exercise reasonable care when selecting an employee to act on their behalf.
    \item[\textbullet] \textbf{Negligent supervision} is a form of primary liability where the principal is held responsible for failing to provide training or supervise an agent who is found to have caused harm to a third party.\footnote{\textit{Restatement (Third) of Agency § 7.03 (2006)}}  For example, an auction house could be held liable for hiring ex-convicts to serve as security guards and and failing to control them when they forcefully removed customers from the premises and harming them in the process.\footnote{\textit{American Auto. Auction, Inc. v. Titsworth}, 730 S.W.2d 499 (1987)}    
    \item[\textbullet] \textbf{Vicarious liability} is a form of secondary liability -- because the principal is responsible for the actions of another person based on the relationship between the principal and the agent -- where the agent caused harm to a third party within the scope of the agency relationship\footnote{\textit{Restatement (Third) of Agency § 7.05 (2006)}}. The scope of the relationship defines the boundaries of the principal's responsibility.  If the principal provides some instructions to the agent and the agents carries them to the letter and then causes harm to a third party, the principal is held vicariously liable and the third party can sue the agent and the principal.  However, even if the agent deviates from the principal's instructions, the principal may still be held vicariously liable.  For example, Company A hires 3 taxi drivers to work in 8-hour shifts in New York City.  Driver X hit Pedestrian Y and broke Pedestrian Y's leg while carrying a fare from Wall Street to Times Square.  Pedestrian sues Driver X and Company A to recover his medical bills and the pain and suffering from having a leg problem.\footnote{\textit{Restatement (Third) of Agency 7.07 Employee Acting Within Scope of Employment}}
\end{enumerate}

\subsection{Hypotheticals}
\label{app:hypothete}
The following examples of liability analysis are written from a legal perspective using hypothetical names. They contrast with those discussed in the main text sections 4-5 using examples largely from machine learning, language models and agents.

\paragraph{Agentic home security system} Acme corp. offers private security system services to independent homeowners.  When hired, the company sends employees to install the security system into a home including components such as cameras, infrared, electric fences, etc.  Aside from installation, Acme relies on an AI agent with the ability to trigger various systems (e.g., automated sprinklers) to monitor homes and intrusions.  In case of an intrusion, Acme's AI agent can contact the police or private security for a welfare check or deploy a drone to conduct a scan of the property. The drone surveillance is provided by comp. developing AI-powered drones. The comp. uses its own system to control the drones and carry out security checks. All those communications are AI to AI and no human comes into the decision loop whether to trigger a sprinkler or send a drone, but some humans may participate in the act during checks (e.g., police officers).  

In 2025, Peter Principal hired Acme corp. to provide security to its chemical factory.  A few months later, Ted Thieve breaks into the factory.  Acme's AI agent detects the intrusion with its infrared sensors installed on the periphery of the factory and asks AI drone to send a drone to provide aerial pictures of the factory to confirm the intrusion.  Before it could even reach Peter Principal's factory.  The drone loses control and drops on Victoria Victim's car, damaging it.  Victoria Victim sues Peter Principal, Acme corp. and AI-powered drone to recover the damages caused to her car.

\paragraph{Agentic package delivery system} ABC corp. offers services for delivery systems.  They claim to be able to deliver any package weighing two pounds or under to any of the 48 contiguous states within 24 hours.  To be able to do so, ABC corp. uses a sophisticated agentic system that decides the route, hires local delivery services and handles the packages.  In other words, its system coordinates with and contract various service providers.  Many of the local service providers use their own agentic system to decide what deliver service to accept, which to reject, the route, and the costing.  The agentic systems do more than logistics.  In many cases, the agentic system also uses delivery robots to deliver the packages and those delivery robots are controlled by those same agentic system.  All those communications are AI to AI and no human comes into the decision loop, but some humans may take part in delivering service (e.g., truck driver).  

In 2025, Peter Principal hired ABC corp. to deliver a new birthday cake from New York City, New York to Salt Lake City, Utah.  ABC's agentic system accepted the package.  It contacted Delivery Express's agentic system to fly the package from New York City to Utah and Drone Delivery to take the package from the airport to the delivery address.  During the flight, the AI-powered drone delivery service has a malfunction and drops the package on Victoria Victim, who is injured.  Victoria Victim sues Peter Principal, ABC corp., Delivery Express, and Drone Delivery to recover the medical bills and the pain and suffering.

\section{Principal-agent analysis of an LLM-related legal case}
\label{sec:mata}
\begin{figure}[htbp!]
    \centering
    \includegraphics[width=0.9\linewidth]{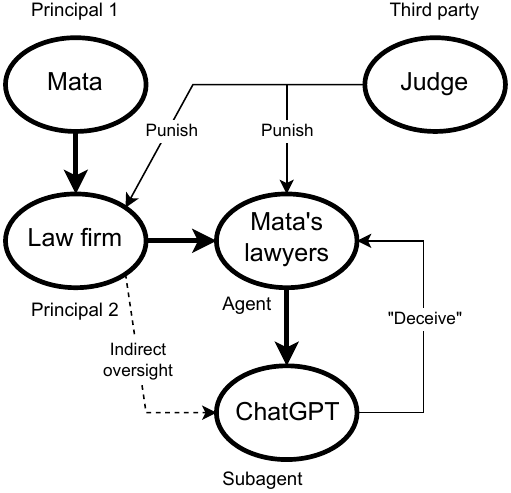}
    \caption{Principal-agent analysis of Mata vs. Avianca, Inc.}
    \label{fig:case_study}
\end{figure}

The case features a few different principal-agent relations, starting with Roberto Mata, the plaintiff. The segment relevant for the current discussion concerns the law firm that hired the two lawyers representing Mata. A part of the tasks that the lawyers do was delegated to the LLM, ChatGPT, which is the subagent of the law firm. The law firm can control the tools that the lawyers use through the employment contract, but ChatGPT was \textit{not} explicitly excluded. The lawyers defended the fake cases generated by ChatGPT and were therefore ruled by the judge to be in bad faith. The law firm also received punishment alongside its employees.
The case of Mata v. Avianca, Inc. (above) was much discussed in the public media in 2023 as an early incident involving the use of LLMs. The delegational structure of major entities involved in the case is (Fig. \ref{fig:case_study})
\begin{align*}
    \text{Roberto Mata} &\rightarrow \text{Law firm} \\
    &\rightarrow \text{Mata's lawyers} \\
    &\rightarrow \text{ChatGPT}
\end{align*}
\begin{tcolorbox}[colback=Gray!30, boxrule=1pt, left=2mm, right=2mm, top=2mm, bottom=2mm, rounded corners, coltitle=White, drop shadow=black!50!white, fonttitle=\bfseries, title=\text{Mata v. Avianca, Inc.}]
\vspace{-0.2em}
\footnotesize
\textit{In 2023, a US Federal Judge reprimanded two lawyers and their law firm for acting in bad faith and making misleading statements to the court.  Their crime?  They trusted ChatGPT.\footnote{\textit{Mata v. Avianca, Inc.}, 678 F. Supp. 3d 443 (SDNY 2023)} ChatGPT made up cases, the lawyers failed to notice, but the judge did. The judge ordered the lawyers to produce the cases. The lawyers ``doubled down and did not begin to dribble out the truth'' for another few weeks. The court punished the lawyers for the actions of ChatGPT and their law firm in the process. ChatGPT (and its provider OpenAI) did not bear any responsibility because the lawyers were the ones who had a duty to provide accurate information to the court. The lawyers responded that: ``We made a good faith mistake in failing to believe that a piece of technology could be making up cases out of whole cloth.'' These lawyers and their principal learned not to trust AIs.}
\vspace{-0.2em}
\end{tcolorbox}

The prompting method that the lawyers used was not explicitly discussed in the court proceeding. ChatGPT is treated as a subagent that carries out the lawyers' task request. It was largely regarded as an instance of LLM hallucination (or confabulation) in the news media rather than deception. The judicial verdict was finalized based on the behavior of the lawyers. The legal case highlights the shortcomings of current legal frameworks and the technical gap to fully comprehend the behaviors of AI systems in the presence of their environment (involving humans) and make use of this evidence in attributing blame.

\bibliography{custom}

\begin{thebibliography}{164}
\providecommand{\natexlab}[1]{#1}

\bibitem[{Abraham(2012)}]{abraham2012forms}
Kenneth~S Abraham. 2012.
\newblock \emph{The Forms and Functions of Tort Law}.
\newblock Foundation Press.

\bibitem[{Abraham and Rabin(2019)}]{abraham2019automated}
Kenneth~S Abraham and Robert~L Rabin. 2019.
\newblock \href {https://www.jstor.org/stable/26842234} {Automated vehicles and manufacturer responsibility for accidents}.
\newblock \emph{Virginia Law Review}, 105(1):127--171.

\bibitem[{Akerlof(1970)}]{akerlof1970market}
George~A Akerlof. 1970.
\newblock \href {https://doi.org/10.2307/1879431} {The market for “lemons”: Quality uncertainty and the market mechanism}.
\newblock \emph{The Quarterly Journal of Economics}, 84:488--500.

\bibitem[{{American Law Institute}(1965)}]{rest_tort_2}
{American Law Institute}. 1965.
\newblock \emph{Restatement (Second) of Torts}.
\newblock American Law Institute Publishers, Philadelphia.

\bibitem[{{American Law Institute}(2006)}]{rest_agency_3}
{American Law Institute}. 2006.
\newblock \emph{Restatement (Third) of Agency}.
\newblock American Law Institute Publishers, Philadelphia.

\bibitem[{{American Law Institute}(2010)}]{rest_tort_3}
{American Law Institute}. 2010.
\newblock \emph{Restatement (Third) of Torts}.
\newblock American Law Institute Publishers, Philadelphia.
\newblock Various volumes, e.g., Liability for Physical and Emotional Harm, Products Liability.

\bibitem[{Anwar et~al.(2024)Anwar, Saparov, Rando, Paleka, Turpin, Hase, Lubana, Jenner, Casper, Sourbut, Edelman, Zhang, Günther, Korinek, Hernandez-Orallo, Hammond, Bigelow, Pan, Langosco, Korbak, Zhang, Zhong, hEigeartaigh, Recchia, Corsi, Chan, Anderljung, Edwards, Petrov, Witt, Motwani, Bengio, Chen, Torr, Albanie, Maharaj, Foerster, Tramèr, He, Kasirzadeh, Choi, and Krueger}]{anwar_foundational_2024}
Usman Anwar, Abulhair Saparov, Javier Rando, Daniel Paleka, Miles Turpin, Peter Hase, Ekdeep~Singh Lubana, Erik Jenner, Stephen Casper, Oliver Sourbut, Benjamin~L. Edelman, Zhaowei Zhang, Mario Günther, Anton Korinek, Jose Hernandez-Orallo, Lewis Hammond, Eric~J. Bigelow, Alexander Pan, Lauro Langosco, and 23 others. 2024.
\newblock \href {https://openreview.net/forum?id=oVTkOs8Pka} {Foundational {Challenges} in {Assuring} {Alignment} and {Safety} of {Large} {Language} {Models}}.
\newblock \emph{Transactions on Machine Learning Research}.

\bibitem[{Arrow(1963)}]{arrow1963uncertainty}
Kenneth~J Arrow. 1963.
\newblock \href {https://www.jstor.org/stable/1812044} {Uncertainty and the welfare economics of medical care}.
\newblock \emph{The American Economic Review}, 53:941--973.

\bibitem[{Athey et~al.(2020)Athey, Bryan, and Gans}]{athey_allocation_2020}
Susan~C. Athey, Kevin~A. Bryan, and Joshua~S. Gans. 2020.
\newblock \href {https://doi.org/10.1257/pandp.20201034} {The {Allocation} of {Decision} {Authority} to {Human} and {Artificial} {Intelligence}}.
\newblock \emph{AEA Papers and Proceedings}, 110:80--84.

\bibitem[{Baird and Maruping(2021)}]{baird_next_2021}
Aaron Baird and Likoebe Maruping. 2021.
\newblock \href {https://aisel.aisnet.org/misq/vol45/iss1/12} {The {Next} {Generation} of {Research} on {IS} {Use}: {A} {Theoretical} {Framework} of {Delegation} to and from {Agentic} {IS} {Artifacts}}.
\newblock \emph{Management Information Systems Quarterly}, 45(1):315--341.

\bibitem[{Balesni et~al.(2024)Balesni, Hobbhahn, Lindner, Meinke, Korbak, Clymer, Shlegeris, Scheurer, Stix, Shah, Goldowsky-Dill, Braun, Chughtai, Evans, Kokotajlo, and Bushnaq}]{balesni_towards_2024}
Mikita Balesni, Marius Hobbhahn, David Lindner, Alexander Meinke, Tomek Korbak, Joshua Clymer, Buck Shlegeris, Jérémy Scheurer, Charlotte Stix, Rusheb Shah, Nicholas Goldowsky-Dill, Dan Braun, Bilal Chughtai, Owain Evans, Daniel Kokotajlo, and Lucius Bushnaq. 2024.
\newblock \href {https://doi.org/10.48550/arXiv.2411.03336} {Towards evaluations-based safety cases for {AI} scheming}.
\newblock \emph{arXiv preprint}.
\newblock ArXiv:2411.03336 [cs].

\bibitem[{Barandiaran and Almendros(2024)}]{barandiaran_transforming_2024}
Xabier~E. Barandiaran and Lola~S. Almendros. 2024.
\newblock \href {https://doi.org/10.48550/arXiv.2407.10735} {Transforming {Agency}. {On} the mode of existence of {Large} {Language} {Models}}.
\newblock \emph{arXiv preprint}.
\newblock ArXiv:2407.10735 [cs].

\bibitem[{Barbi et~al.(2025)Barbi, Yoran, and Geva}]{barbi_preventing_2025}
Ohav Barbi, Ori Yoran, and Mor Geva. 2025.
\newblock \href {https://doi.org/10.48550/arXiv.2502.05986} {Preventing {Rogue} {Agents} {Improves} {Multi}-{Agent} {Collaboration}}.
\newblock \emph{arXiv preprint}.
\newblock ArXiv:2502.05986 [cs].

\bibitem[{Barfield(2018)}]{barfield2018liability}
Woodrow Barfield. 2018.
\newblock Liability for autonomous and artificially intelligent robots.
\newblock \emph{Paladyn, Journal of Behavioral Robotics}, 9(1):193--203.

\bibitem[{Bass et~al.(2021)Bass, Clements, and Kazman}]{bass_software_2021}
Len Bass, Paul Clements, and Rick Kazman. 2021.
\newblock \emph{Software {Architecture} in {Practice}}.
\newblock Addison-Wesley Professional.

\bibitem[{Bhatt et~al.(2025)Bhatt, Kapoor, Upadhyay, Sucholutsky, Quinzan, Collins, Weller, Wilson, and Zafar}]{bhatt_when_2025}
Umang Bhatt, Sanyam Kapoor, Mihir Upadhyay, Ilia Sucholutsky, Francesco Quinzan, Katherine~M. Collins, Adrian Weller, Andrew~Gordon Wilson, and Muhammad~Bilal Zafar. 2025.
\newblock \href {https://doi.org/10.48550/arXiv.2503.13577} {When {Should} {We} {Orchestrate} {Multiple} {Agents}?}
\newblock \emph{arXiv preprint}.
\newblock ArXiv:2503.13577 [cs].

\bibitem[{Binz and Schulz(2023)}]{binz_turning_2023}
Marcel Binz and Eric Schulz. 2023.
\newblock \href {https://openreview.net/forum?id=eiC4BKypf1} {Turning large language models into cognitive models}.
\newblock In \emph{The Twelfth International Conference on Learning Representations}.

\bibitem[{Blonz(2023)}]{blonz_costs_2023}
Joshua~A. Blonz. 2023.
\newblock \href {https://doi.org/10.1257/pol.20210208} {The {Costs} of {Misaligned} {Incentives}: {Energy} {Inefficiency} and the {Principal}-{Agent} {Problem}}.
\newblock \emph{American Economic Journal: Economic Policy}, 15(3):286--321.

\bibitem[{Bourtoule et~al.(2021)Bourtoule, Chandrasekaran, Choquette-Choo, Jia, Travers, Zhang, Lie, and Papernot}]{bourtoule2021machine}
Lucas Bourtoule, Varun Chandrasekaran, Christopher~A Choquette-Choo, Hengrui Jia, Adelin Travers, Baiwu Zhang, David Lie, and Nicolas Papernot. 2021.
\newblock \href {https://doi.org/10.1109/SP40001.2021.00019} {Machine unlearning}.
\newblock In \emph{2021 IEEE symposium on security and privacy (SP)}, pages 141--159. IEEE.

\bibitem[{Bousetouane(2025)}]{bousetouane_agentic_2025}
Fouad Bousetouane. 2025.
\newblock \href {https://doi.org/10.48550/arXiv.2501.00881} {Agentic {Systems}: {A} {Guide} to {Transforming} {Industries} with {Vertical} {AI} {Agents}}.
\newblock \emph{arXiv preprint}.
\newblock ArXiv:2501.00881 [cs].

\bibitem[{Buiten et~al.(2023)Buiten, de~Streel, and Peitz}]{buiten_law_2023}
Miriam Buiten, Alexandre de~Streel, and Martin Peitz. 2023.
\newblock \href {https://doi.org/10.1016/j.clsr.2023.105794} {The law and economics of {AI} liability}.
\newblock \emph{Computer Law \& Security Review}, 48:105794.

\bibitem[{Buiten(2024)}]{buiten_product_2024}
Miriam~C. Buiten. 2024.
\newblock \href {https://doi.org/10.1007/s10657-024-09794-z} {Product liability for defective {AI}}.
\newblock \emph{European Journal of Law and Economics}, 57(1):239--273.

\bibitem[{Burtell and Woodside(2023)}]{burtell2023artificial}
Matthew Burtell and Thomas Woodside. 2023.
\newblock \href {https://doi.org/10.48550/arXiv.2303.08721} {Artificial influence: An analysis of ai-driven persuasion}.
\newblock \emph{arXiv preprint arXiv:2303.08721}.

\bibitem[{Butlin(2025)}]{butlin_agency_2025}
Patrick Butlin. 2025.
\newblock \href {https://doi.org/10.1080/0020174X.2024.2439995} {The agency in language agents}.
\newblock \emph{Inquiry}, 0(0):1--21.
\newblock Publisher: Routledge \_eprint: https://doi.org/10.1080/0020174X.2024.2439995.

\bibitem[{Calabresi and Melamed(1972)}]{calabresi1972property}
Guido Calabresi and A~Douglas Melamed. 1972.
\newblock \href {https://doi.org/10.2307/1340059} {Property rules, liability rules, and inalienability: one view of the cathedral}.
\newblock \emph{Havard Law Review}, 85:1089--1128.

\bibitem[{Camacho(1993)}]{camacho1993avoid}
Rodolfo~A Camacho. 1993.
\newblock \href {https://heinonline.org/HOL/P?h=hein.journals/whitlr14&i=799} {How to avoid negligent hiring litigation}.
\newblock \emph{Whittier Law Review}, 14:787--808.

\bibitem[{Campbell and Wu(2011)}]{campbell_mas_2011}
Adam Campbell and Annie~S. Wu. 2011.
\newblock \href {https://doi.org/10.1007/s10458-010-9127-4} {Multi-agent role allocation: issues, approaches, and multiple perspectives}.
\newblock \emph{Autonomous Agents and Multi-Agent Systems}, 22(2):317--355.

\bibitem[{Campedelli et~al.(2024)Campedelli, Penzo, Stefan, Dess{\`\i}, Guerini, Lepri, and Staiano}]{campedelli2024want}
Gian~Maria Campedelli, Nicol{\`o} Penzo, Massimo Stefan, Roberto Dess{\`\i}, Marco Guerini, Bruno Lepri, and Jacopo Staiano. 2024.
\newblock \href {https://doi.org/10.48550/arXiv.2410.07109} {I want to break free! persuasion and anti-social behavior of llms in multi-agent settings with social hierarchy}.
\newblock \emph{arXiv preprint arXiv:2410.07109}.

\bibitem[{Carbonara et~al.(2016)Carbonara, Guerra, and Parisi}]{carbonara2016sharing}
Emanuela Carbonara, Alice Guerra, and Francesco Parisi. 2016.
\newblock \href {https://doi.org/10.1086/685498} {Sharing residual liability: the cheapest cost avoider revisited}.
\newblock \emph{The Journal of Legal Studies}, 45(1):173--201.

\bibitem[{Carroll et~al.(2023)Carroll, Chan, Ashton, and Krueger}]{carroll_characterizing_2023}
Micah Carroll, Alan Chan, Henry Ashton, and David Krueger. 2023.
\newblock \href {https://doi.org/10.1145/3617694.3623226} {Characterizing {Manipulation} from {AI} {Systems}}.
\newblock In \emph{Equity and {Access} in {Algorithms}, {Mechanisms}, and {Optimization}}, pages 1--13, New York, NY, USA. ACM.

\bibitem[{Casey(2019)}]{casey2019robot}
Bryan Casey. 2019.
\newblock \href {https://heinonline.org/HOL/P?h=hein.journals/glj108&i=231} {Robot {Ipsa} {Loquitur}}.
\newblock \emph{Georgetown Law Journal}, 108(2):225--286.

\bibitem[{Castelfranchi and Falcone(1998)}]{castelfranchi_towards_1998}
Cristiano Castelfranchi and Rino Falcone. 1998.
\newblock \href {https://doi.org/10.1016/S0921-8890(98)00028-1} {Towards a theory of delegation for agent-based systems}.
\newblock \emph{Robotics and Autonomous Systems}, 24(3):141--157.

\bibitem[{Cavico et~al.(2016)Cavico, Mujtaba, Samuel, and Muffler}]{cavico2016tort}
Frank~J Cavico, Bahaudin~G Mujtaba, Marissa Samuel, and Stephen~C Muffler. 2016.
\newblock \href {https://nsuworks.nova.edu/hcbe_facarticles/559/} {The tort of negligence in employment hiring, supervision, and retention}.
\newblock \emph{American Journal of Business and Society}, 1(4):205.

\bibitem[{Chan et~al.(2024)Chan, Ezell, Kaufmann, Wei, Hammond, Bradley, Bluemke, Rajkumar, Krueger, Kolt, Heim, and Anderljung}]{chan_visibility_2024}
Alan Chan, Carson Ezell, Max Kaufmann, Kevin Wei, Lewis Hammond, Herbie Bradley, Emma Bluemke, Nitarshan Rajkumar, David Krueger, Noam Kolt, Lennart Heim, and Markus Anderljung. 2024.
\newblock \href {https://doi.org/10.1145/3630106.3658948} {Visibility into {AI} {Agents}}.
\newblock In \emph{Proceedings of the 2024 {ACM} {Conference} on {Fairness}, {Accountability}, and {Transparency}}, {FAccT} '24, pages 958--973, New York, NY, USA. Association for Computing Machinery.

\bibitem[{Chan et~al.(2023)Chan, Salganik, Markelius, Pang, Rajkumar, Krasheninnikov, Langosco, He, Duan, Carroll, Lin, Mayhew, Collins, Molamohammadi, Burden, Zhao, Rismani, Voudouris, Bhatt, Weller, Krueger, and Maharaj}]{chan_harms_2023}
Alan Chan, Rebecca Salganik, Alva Markelius, Chris Pang, Nitarshan Rajkumar, Dmitrii Krasheninnikov, Lauro Langosco, Zhonghao He, Yawen Duan, Micah Carroll, Michelle Lin, Alex Mayhew, Katherine Collins, Maryam Molamohammadi, John Burden, Wanru Zhao, Shalaleh Rismani, Konstantinos Voudouris, Umang Bhatt, and 3 others. 2023.
\newblock \href {https://doi.org/10.1145/3593013.3594033} {Harms from {Increasingly} {Agentic} {Algorithmic} {Systems}}.
\newblock In \emph{Proceedings of the 2023 {ACM} {Conference} on {Fairness}, {Accountability}, and {Transparency}}, {FAccT} '23, pages 651--666, New York, NY, USA. Association for Computing Machinery.

\bibitem[{Chan et~al.(2025)Chan, Wei, Huang, Rajkumar, Perrier, Lazar, Hadfield, and Anderljung}]{chan_infrastructure_2025}
Alan Chan, Kevin Wei, Sihao Huang, Nitarshan Rajkumar, Elija Perrier, Seth Lazar, Gillian~K. Hadfield, and Markus Anderljung. 2025.
\newblock \href {https://doi.org/10.48550/arXiv.2501.10114} {Infrastructure for {AI} {Agents}}.
\newblock \emph{arXiv preprint}.
\newblock ArXiv:2501.10114 [cs].

\bibitem[{Chen et~al.(2024{\natexlab{a}})Chen, Wang, Xu, Yuan, Zhang, Shi, Xie, Li, Yang, Zhu, Chen, Li, Chen, Hu, Wu, Ren, Fu, and Xiao}]{chen_persona_2024}
Jiangjie Chen, Xintao Wang, Rui Xu, Siyu Yuan, Yikai Zhang, Wei Shi, Jian Xie, Shuang Li, Ruihan Yang, Tinghui Zhu, Aili Chen, Nianqi Li, Lida Chen, Caiyu Hu, Siye Wu, Scott Ren, Ziquan Fu, and Yanghua Xiao. 2024{\natexlab{a}}.
\newblock \href {https://openreview.net/forum?id=xrO70E8UIZ} {From {Persona} to {Personalization}: {A} {Survey} on {Role}-{Playing} {Language} {Agents}}.
\newblock \emph{Transactions on Machine Learning Research}.

\bibitem[{Chen et~al.(2024{\natexlab{b}})Chen, Pesaranghader, Sadhu, and Yi}]{chen_can_2024}
Yanan Chen, Ali Pesaranghader, Tanmana Sadhu, and Dong~Hoon Yi. 2024{\natexlab{b}}.
\newblock \href {https://doi.org/10.48550/arXiv.2408.06318} {Can {We} {Rely} on {LLM} {Agents} to {Draft} {Long}-{Horizon} {Plans}? {Let}'s {Take} {TravelPlanner} as an {Example}}.
\newblock \emph{arXiv preprint}.
\newblock ArXiv:2408.06318 [cs].

\bibitem[{Choi(2020)}]{choi_software_2020}
Bryan~H. Choi. 2020.
\newblock \href {https://heinonline.org/HOL/P?h=hein.journals/hjlt33&i=569} {Software as a {Profession}}.
\newblock \emph{Harvard Journal of Law \& Technology (Harvard JOLT)}, 33(2):557--638.

\bibitem[{Chopra and White(2011)}]{chopra_legal_2011}
Samir Chopra and Laurence~F. White. 2011.
\newblock \emph{A {Legal} {Theory} for {Autonomous} {Artificial} {Agents}}.
\newblock University of Michigan Press, Ann Arbor.

\bibitem[{Cihon(2024)}]{cihon_chilling_2024}
Peter Cihon. 2024.
\newblock \href {https://blog.genlaw.org/pdfs/genlaw_icml2024/79.pdf} {Chilling autonomy: {Policy} enforcement for human oversight of {AI} agents}.
\newblock In \emph{41st {International} {Conference} on {Machine} {Learning}, {Workshop} on {Generative} {AI} and {Law}}.

\bibitem[{Coase(1960)}]{coase1960problem}
Ronald~Harry Coase. 1960.
\newblock \href {https://www.jstor.org/stable/724810} {The problem of social cost}.
\newblock \emph{The Journal of Law and Economics}, 3:1--44.

\bibitem[{Cohen et~al.(2024)Cohen, Kolt, Bengio, Hadfield, and Russell}]{cohen_regulating_2024}
Michael~K. Cohen, Noam Kolt, Yoshua Bengio, Gillian~K. Hadfield, and Stuart Russell. 2024.
\newblock \href {https://doi.org/10.1126/science.adl0625} {Regulating advanced artificial agents}.
\newblock \emph{Science}, 384(6691):36--38.
\newblock Publisher: American Association for the Advancement of Science.

\bibitem[{Cooper et~al.(2024)Cooper, Choquette-Choo, Bogen, Jagielski, Filippova, Liu, Chouldechova, Hayes, Huang, Mireshghallah, Shumailov, Triantafillou, Kairouz, Mitchell, Liang, Ho, Choi, Koyejo, Delgado, Grimmelmann, Shmatikov, Sa, Barocas, Cyphert, Lemley, boyd, Vaughan, Brundage, Bau, Neel, Jacobs, Terzis, Wallach, Papernot, and Lee}]{cooper_machine_2024}
A.~Feder Cooper, Christopher~A. Choquette-Choo, Miranda Bogen, Matthew Jagielski, Katja Filippova, Ken~Ziyu Liu, Alexandra Chouldechova, Jamie Hayes, Yangsibo Huang, Niloofar Mireshghallah, Ilia Shumailov, Eleni Triantafillou, Peter Kairouz, Nicole Mitchell, Percy Liang, Daniel~E. Ho, Yejin Choi, Sanmi Koyejo, Fernando Delgado, and 16 others. 2024.
\newblock \href {https://doi.org/10.48550/arXiv.2412.06966} {Machine {Unlearning} {Doesn}'t {Do} {What} {You} {Think}: {Lessons} for {Generative} {AI} {Policy}, {Research}, and {Practice}}.
\newblock \emph{arXiv preprint}.
\newblock ArXiv:2412.06966 [cs].

\bibitem[{Criado et~al.(2011)Criado, Argente, and Botti}]{criado_open_2011}
N.~Criado, E.~Argente, and V.~Botti. 2011.
\newblock \href {https://doi.org/10.3233/AIC-2011-0502} {Open issues for normative multi-agent systems}.
\newblock \emph{AI Communications}, 24(3):233--264.
\newblock Publisher: IOS Press.

\bibitem[{Critch and Krueger(2020)}]{critch_xrisk_2020}
Andrew Critch and David Krueger. 2020.
\newblock \href {https://doi.org/10.48550/arXiv.2006.04948} {{AI} {Research} {Considerations} for {Human} {Existential} {Safety} ({ARCHES})}.
\newblock \emph{arXiv preprint}.
\newblock ArXiv:2006.04948 [cs].

\bibitem[{Custers et~al.(2025)Custers, Lahmann, and Scott}]{custers2025liability}
Bart Custers, Henning Lahmann, and Benjamyn~I Scott. 2025.
\newblock \href {https://doi.org/10.1007/s00146-024-02137-1} {From liability gaps to liability overlaps: shared responsibilities and fiduciary duties in ai and other complex technologies}.
\newblock \emph{AI \& SOCIETY}, pages 1--16.

\bibitem[{Dai(2024)}]{dai_position_2024}
Jessica Dai. 2024.
\newblock \href {https://openreview.net/forum?id=4XlGXIh2BB} {Position: {Beyond} {Personhood}: {Agency}, {Accountability}, and the {Limits} of {Anthropomorphic} {Ethical} {Analysis}}.
\newblock In \emph{Forty-first International Conference on Machine Learning}.

\bibitem[{Das(2025)}]{das_agency_2025}
Parashar Das. 2025.
\newblock \href {https://doi.org/10.48550/arXiv.2502.10434} {Agency in {Artificial} {Intelligence} {Systems}}.
\newblock \emph{arXiv preprint}.
\newblock ArXiv:2502.10434 [cs].

\bibitem[{Diamantis(2023)}]{diamantis_vicarious_2023}
Mihailis~E. Diamantis. 2023.
\newblock \href {https://heinonline.org/HOL/P?h=hein.journals/indana99&i=330} {Vicarious {Liability} for {AI}}.
\newblock \emph{Indiana Law Journal}, 99(1):317--334.

\bibitem[{Diaz et~al.(2024)Diaz, Leibo, and Paull}]{diaz_milnor-myerson_2024}
Manfred Diaz, Joel~Z. Leibo, and Liam Paull. 2024.
\newblock \href {https://openreview.net/forum?id=LNYJrXdk45} {Milnor-{Myerson} {Games} and {The} {Principles} of {Artificial} {Principal}-{Agent} {Problems}}.
\newblock In \emph{Finding the Frame: An RLC Workshop for Examining Conceptual Frameworks}.

\bibitem[{Dung(2024)}]{dung_understanding_2024}
Leonard Dung. 2024.
\newblock \href {https://doi.org/10.1093/pq/pqae010} {Understanding {Artificial} {Agency}}.
\newblock \emph{The Philosophical Quarterly}, page pqae010.

\bibitem[{Eisenhardt(1989)}]{eisenhardt_agency_1989}
Kathleen~M. Eisenhardt. 1989.
\newblock \href {https://doi.org/10.2307/258191} {Agency {Theory}: {An} {Assessment} and {Review}}.
\newblock \emph{The Academy of Management Review}, 14(1):57--74.
\newblock Publisher: Academy of Management.

\bibitem[{Everitt et~al.(2021)Everitt, Carey, Langlois, Ortega, and Legg}]{everitt_agent_2021}
Tom Everitt, Ryan Carey, Eric~D. Langlois, Pedro~A. Ortega, and Shane Legg. 2021.
\newblock \href {https://doi.org/10.1609/aaai.v35i13.17368} {Agent {Incentives}: {A} {Causal} {Perspective}}.
\newblock \emph{Proceedings of the AAAI Conference on Artificial Intelligence}, 35(13):11487--11495.
\newblock Number: 13.

\bibitem[{Fama(1980)}]{fama_agency_1980}
Eugene~F. Fama. 1980.
\newblock \href {https://doi.org/10.1086/260866} {Agency {Problems} and the {Theory} of the {Firm}}.
\newblock \emph{Journal of Political Economy}, 88(2):288--307.
\newblock Publisher: The University of Chicago Press.

\bibitem[{Fama and Jensen(1983)}]{fama_agency_1983}
Eugene~F. Fama and Michael~C. Jensen. 1983.
\newblock \href {https://www.jstor.org/stable/725105} {Agency {Problems} and {Residual} {Claims}}.
\newblock \emph{The Journal of Law \& Economics}, 26(2):327--349.
\newblock Publisher: [University of Chicago Press, Booth School of Business, University of Chicago, University of Chicago Law School].

\bibitem[{Fish et~al.(2024)Fish, Gonczarowski, and Shorrer}]{fish_algorithmic_2024}
Sara Fish, Yannai~A. Gonczarowski, and Ran~I. Shorrer. 2024.
\newblock \href {https://doi.org/10.48550/arXiv.2404.00806} {Algorithmic {Collusion} by {Large} {Language} {Models}}.
\newblock \emph{arXiv preprint}.
\newblock ArXiv:2404.00806 [econ].

\bibitem[{Foxabbott et~al.(2023)Foxabbott, Deverett, Senft, Dower, and Hammond}]{foxabbott_defining_2023}
Jack Foxabbott, Sam Deverett, Kaspar Senft, Samuel Dower, and Lewis Hammond. 2023.
\newblock \href {https://openreview.net/forum?id=tF464LogjS} {Defining and {Mitigating} {Collusion} in {Multi}-{Agent} {Systems}}.

\bibitem[{Fraser et~al.(2022)Fraser, Simcock, and Snoswell}]{fraser2022ai}
Henry Fraser, Rhyle Simcock, and Aaron~J Snoswell. 2022.
\newblock \href {https://dl.acm.org/doi/pdf/10.1145/3531146.3533084} {Ai opacity and explainability in tort litigation}.
\newblock In \emph{Proceedings of the 2022 ACM Conference on Fairness, Accountability, and Transparency}, pages 185--196.

\bibitem[{Furr(2024)}]{Furr_2024}
Jessica Furr. 2024.
\newblock \href {https://www.thelawverse.com/p/the-untouchables-why-software-companies#§ii-is-software-a-service-or-a-product} {The untouchables: Why software companies escape liability for faulty software?}
\newblock \emph{The LawVerse Substack}.

\bibitem[{Gabison and Buiten(2020)}]{gabison_platform_2020}
Garry~A. Gabison and Miriam~C. Buiten. 2020.
\newblock \href {https://heinonline.org/HOL/P?h=hein.journals/cstlr21&i=244} {Platform {Liability} in {Copyright} {Enforcement}}.
\newblock \emph{Columbia Science and Technology Law Review}, 21(2):237--281.

\bibitem[{Gan et~al.(2024)Gan, Han, Wu, and Xu}]{gan_generalized_2024}
Jiarui Gan, Minbiao Han, Jibang Wu, and Haifeng Xu. 2024.
\newblock \href {https://doi.org/10.48550/arXiv.2209.01146} {Generalized {Principal}-{Agency}: {Contracts}, {Information}, {Games} and {Beyond}}.
\newblock \emph{arXiv preprint}.
\newblock ArXiv:2209.01146 [cs].

\bibitem[{Geiger et~al.(2024)Geiger, Ibeling, Zur, Chaudhary, Chauhan, Huang, Arora, Wu, Goodman, Potts, and Icard}]{geiger_causal_2024}
Atticus Geiger, Duligur Ibeling, Amir Zur, Maheep Chaudhary, Sonakshi Chauhan, Jing Huang, Aryaman Arora, Zhengxuan Wu, Noah Goodman, Christopher Potts, and Thomas Icard. 2024.
\newblock \href {https://doi.org/10.48550/arXiv.2301.04709} {Causal {Abstraction}: {A} {Theoretical} {Foundation} for {Mechanistic} {Interpretability}}.
\newblock \emph{arXiv preprint}.
\newblock ArXiv:2301.04709 [cs].

\bibitem[{Geistfeld et~al.(2022)Geistfeld, Karner, and Koch}]{geistfeld_comparative_2022}
Mark~A. Geistfeld, Ernst Karner, and Bernhard~A. Koch. 2022.
\newblock \href {https://doi.org/10.1515/9783110775402-001} {Comparative {Law} {Study} on {Civil} {Liability} for {Artificial} {Intelligence}}.
\newblock In Ernst Karner, Bernhard~A. Koch, Mark~A. Geistfeld, and Christiane Wendehorst, editors, \emph{Civil {Liability} for {Artificial} {Intelligence} and {Software}}, pages 1--184. De Gruyter.

\bibitem[{Gemignani(1980)}]{gemignani1980product}
Michael~C. Gemignani. 1980.
\newblock \href {https://heinonline.org/HOL/P?h=hein.journals/rutcomt8&i=179} {Product {Liability} and {Software}}.
\newblock \emph{Rutgers Computer \& Technology Law Journal}, 8(2):173--204.

\bibitem[{Goldowsky-Dill et~al.(2025)Goldowsky-Dill, Chughtai, Heimersheim, and Hobbhahn}]{gdill_detecting_2025}
Nicholas Goldowsky-Dill, Bilal Chughtai, Stefan Heimersheim, and Marius Hobbhahn. 2025.
\newblock \href {https://doi.org/10.48550/arXiv.2502.03407} {Detecting {Strategic} {Deception} {Using} {Linear} {Probes}}.
\newblock \emph{arXiv preprint}.
\newblock ArXiv:2502.03407 [cs].

\bibitem[{Greenblatt et~al.(2024)Greenblatt, Denison, Wright, Roger, MacDiarmid, Marks, Treutlein, Belonax, Chen, Duvenaud, Khan, Michael, Mindermann, Perez, Petrini, Uesato, Kaplan, Shlegeris, Bowman, and Hubinger}]{greenblatt_alignment_2024}
Ryan Greenblatt, Carson Denison, Benjamin Wright, Fabien Roger, Monte MacDiarmid, Sam Marks, Johannes Treutlein, Tim Belonax, Jack Chen, David Duvenaud, Akbir Khan, Julian Michael, Sören Mindermann, Ethan Perez, Linda Petrini, Jonathan Uesato, Jared Kaplan, Buck Shlegeris, Samuel~R. Bowman, and Evan Hubinger. 2024.
\newblock \href {https://doi.org/10.48550/arXiv.2412.14093} {Alignment faking in large language models}.
\newblock \emph{arXiv preprint}.
\newblock ArXiv:2412.14093 [cs].

\bibitem[{Guo et~al.(2024)Guo, Chen, Wang, Chang, Pei, Chawla, Wiest, and Zhang}]{guo_large_2024}
Taicheng Guo, Xiuying Chen, Yaqi Wang, Ruidi Chang, Shichao Pei, Nitesh~V. Chawla, Olaf Wiest, and Xiangliang Zhang. 2024.
\newblock \href {https://doi.org/10.24963/ijcai.2024/890} {Large {Language} {Model} {Based} {Multi}-agents: {A} {Survey} of {Progress} and {Challenges}}.
\newblock volume~9, pages 8048--8057.
\newblock ISSN: 1045-0823.

\bibitem[{Hadfield-Menell et~al.(2019)Hadfield-Menell, Andrus, and Hadfield}]{hadfield-menell_legible_2019}
Dylan Hadfield-Menell, Mckane Andrus, and Gillian Hadfield. 2019.
\newblock \href {https://doi.org/10.1145/3306618.3314258} {Legible {Normativity} for {AI} {Alignment}: {The} {Value} of {Silly} {Rules}}.
\newblock In \emph{Proceedings of the 2019 {AAAI}/{ACM} {Conference} on {AI}, {Ethics}, and {Society}}, {AIES} '19, pages 115--121, New York, NY, USA. Association for Computing Machinery.

\bibitem[{Hadfield-Menell and Hadfield(2019)}]{hadfield2019incomplete}
Dylan Hadfield-Menell and Gillian~K. Hadfield. 2019.
\newblock \href {https://doi.org/10.1145/3306618.3314250} {Incomplete contracting and ai alignment}.
\newblock In \emph{Proceedings of the 2019 AAAI/ACM Conference on AI, Ethics, and Society}, AIES '19, page 417–422, New York, NY, USA. Association for Computing Machinery.

\bibitem[{Hammond et~al.(2025)Hammond, Chan, Clifton, Hoelscher-Obermaier, Khan, McLean, Smith, Barfuss, Foerster, Gavenčiak, Han, Hughes, Kovařík, Kulveit, Leibo, Oesterheld, Witt, Shah, Wellman, Bova, Cimpeanu, Ezell, Feuillade-Montixi, Franklin, Kran, Krawczuk, Lamparth, Lauffer, Meinke, Motwani, Reuel, Conitzer, Dennis, Gabriel, Gleave, Hadfield, Haghtalab, Kasirzadeh, Krier, Larson, Lehman, Parkes, Piliouras, and Rahwan}]{hammond_multiagent_2025}
Lewis Hammond, Alan Chan, Jesse Clifton, Jason Hoelscher-Obermaier, Akbir Khan, Euan McLean, Chandler Smith, Wolfram Barfuss, Jakob Foerster, Tomáš Gavenčiak, The~Anh Han, Edward Hughes, Vojtěch Kovařík, Jan Kulveit, Joel~Z. Leibo, Caspar Oesterheld, Christian Schroeder~de Witt, Nisarg Shah, Michael Wellman, and 25 others. 2025.
\newblock \href {https://doi.org/10.48550/arXiv.2502.14143} {Multi-{Agent} {Risks} from {Advanced} {AI}}.
\newblock \emph{arXiv preprint}.
\newblock ArXiv:2502.14143 [cs].

\bibitem[{He et~al.(2025)He, Lin, Dong, Xu, Xing, and Liu}]{he_redteaming_2025}
Pengfei He, Yupin Lin, Shen Dong, Han Xu, Yue Xing, and Hui Liu. 2025.
\newblock \href {https://doi.org/10.48550/arXiv.2502.14847} {Red-{Teaming} {LLM} {Multi}-{Agent} {Systems} via {Communication} {Attacks}}.
\newblock \emph{arXiv preprint}.
\newblock ArXiv:2502.14847 [cs].

\bibitem[{He et~al.(2024)He, Wang, Rong, Cheng, and Chen}]{he_security_2024}
Yifeng He, Ethan Wang, Yuyang Rong, Zifei Cheng, and Hao Chen. 2024.
\newblock \href {https://doi.org/10.48550/arXiv.2406.08689} {Security of {AI} {Agents}}.
\newblock \emph{arXiv preprint}.
\newblock ArXiv:2406.08689 [cs].

\bibitem[{Hendrycks(2023)}]{hendrycks_natural_2023}
Dan Hendrycks. 2023.
\newblock \href {https://doi.org/10.48550/arXiv.2303.16200} {Natural {Selection} {Favors} {AIs} over {Humans}}.
\newblock \emph{arXiv preprint}.
\newblock ArXiv:2303.16200 [cs].

\bibitem[{Holmstr{\"o}m(1979)}]{holmstrom1979moral}
Bengt Holmstr{\"o}m. 1979.
\newblock \href {https://www.jstor.org/stable/3003320} {Moral hazard and observability}.
\newblock \emph{The Bell journal of economics}, 10:74--91.

\bibitem[{Hua et~al.(2024)Hua, Yang, Jin, Li, Cheng, Tang, and Zhang}]{hua_trustagent_2024}
Wenyue Hua, Xianjun Yang, Mingyu Jin, Zelong Li, Wei Cheng, Ruixiang Tang, and Yongfeng Zhang. 2024.
\newblock \href {https://doi.org/10.18653/v1/2024.findings-emnlp.585} {{TrustAgent}: {Towards} {Safe} and {Trustworthy} {LLM}-based {Agents}}.
\newblock In \emph{Findings of the {Association} for {Computational} {Linguistics}: {EMNLP} 2024}, pages 10000--10016, Miami, Florida, USA. Association for Computational Linguistics.

\bibitem[{Ivanov et~al.(2024)Ivanov, Dütting, Talgam-Cohen, Wang, and Parkes}]{ivanov_pap_2024}
Dima Ivanov, Paul Dütting, Inbal Talgam-Cohen, Tonghan Wang, and David~C. Parkes. 2024.
\newblock \href {https://doi.org/10.48550/arXiv.2407.18074} {Principal-{Agent} {Reinforcement} {Learning}: {Orchestrating} {AI} {Agents} with {Contracts}}.
\newblock \emph{arXiv preprint}.
\newblock ArXiv:2407.18074 [cs].

\bibitem[{Jain et~al.(2023)Jain, Hitzig, and Mishkin}]{jain_contextual_2023}
Shrey Jain, Zoë Hitzig, and Pamela Mishkin. 2023.
\newblock \href {https://doi.org/10.48550/arXiv.2311.01193} {Contextual {Confidence} and {Generative} {AI}}.
\newblock \emph{arXiv preprint}.
\newblock ArXiv:2311.01193 [cs].

\bibitem[{Jennings(2001)}]{jennings_absoft_2001}
Nicholas~R. Jennings. 2001.
\newblock \href {https://doi.org/10.1145/367211.367250} {An agent-based approach for building complex software systems}.
\newblock \emph{Commun. ACM}, 44(4):35--41.

\bibitem[{Jensen and Meckling(1976)}]{jensen_theory_1976}
Michael~C. Jensen and William~H. Meckling. 1976.
\newblock \href {https://doi.org/10.1016/0304-405X(76)90026-X} {Theory of the firm: {Managerial} behavior, agency costs and ownership structure}.
\newblock \emph{Journal of Financial Economics}, 3(4):305--360.

\bibitem[{Jiang et~al.(2024)Jiang, Ju, Cohen, Mitts, Foss, Kao, Li, and Tian}]{jiang_towards_2024}
Song Jiang, Da~Ju, Andrew Cohen, Sasha Mitts, Aaron Foss, Justine~T. Kao, Xian Li, and Yuandong Tian. 2024.
\newblock \href {https://openreview.net/forum?id=YToS2XetSJ} {Towards {Full} {Delegation}: {Designing} {Ideal} {Agentic} {Behaviors} for {Travel} {Planning}}.
\newblock In \emph{NeurIPS 2024 Workshop on Adaptive Foundation Models}.

\bibitem[{John et~al.(2025)John, Ron F.~Del, Evgeniy, Helen, Idan, Kayla, Ken, Peter, Rakshith, Ron, Tamir~Ishay, Vinnie, Volkan, Allie, Anshuman, Akram, Emmanuel, Eric, Itsik, Kellen, Keren, Kreshnik, Manish~Kumar, Matt, Mohit, Nate, Patrik, Peter, Riggs, Rock, Sahana, Sandy, Srinivas, Subaru, Trent, Victor, Alejandro, Apostol, Chris, Hyrum, Scott, Steve, Vasilios, John, Ron F.~Del, Evgeniy, Helen, Idan, Kayla, Ken, Peter, Rakshith, Ron, Tamir~Ishay, Vinnie, Volkan, Allie, Anshuman, Akram, Emmanuel, Eric, Itsik, Kellen, Keren, Kreshnik, Manish~Kumar, Matt, Mohit, Nate, Patrik, Peter, Riggs, Rock, Sahana, Sandy, Srinivas, Subaru, Trent, Victor, Alejandro, Apostol, Chris, Hyrum, Scott, Steve, and Vasilios}]{john_owasp_2025}
Sotiropoulos John, Rosario Ron F.~Del, Kokuykin Evgeniy, Oakley Helen, Habler Idan, Underkoffler Kayla, Huang Ken, Steffensen Peter, Aralimatti Rakshith, Bitton Ron, Sharbat Tamir~Ishay, Giarrusso Vinnie, Kutal Volkan, Howe Allie, Bhartiya Anshuman, Sheriff Akram, Guilherme Emmanuel, Rogers Eric, Martin Itsik, and 67 others. 2025.
\newblock \href {https://hal.science/hal-04985337} {{OWASP} {Top} 10 for {LLM} {Apps} \& {Gen} {AI} {Agentic} {Security} {Initiative}}.
\newblock Technical report, OWASP.

\bibitem[{Kambhampati et~al.(2024)Kambhampati, Valmeekam, Guan, Verma, Stechly, Bhambri, Saldyt, and Murthy}]{kambhampati_position_2024}
Subbarao Kambhampati, Karthik Valmeekam, Lin Guan, Mudit Verma, Kaya Stechly, Siddhant Bhambri, Lucas~Paul Saldyt, and Anil~B. Murthy. 2024.
\newblock \href {https://openreview.net/forum?id=Th8JPEmH4z} {Position: {LLMs} {Can}’t {Plan}, {But} {Can} {Help} {Planning} in {LLM}-{Modulo} {Frameworks}}.
\newblock In \emph{Forty-first International Conference on Machine Learning}.

\bibitem[{Kampik et~al.(2022)Kampik, Mansour, Boissier, Kirrane, Padget, Payne, Singh, Tamma, and Zimmermann}]{kampik_governance_2022}
Timotheus Kampik, Adnane Mansour, Olivier Boissier, Sabrina Kirrane, Julian Padget, Terry~R. Payne, Munindar~P. Singh, Valentina Tamma, and Antoine Zimmermann. 2022.
\newblock \href {https://doi.org/10.1145/3507910} {Governance of {Autonomous} {Agents} on the {Web}: {Challenges} and {Opportunities}}.
\newblock \emph{ACM Trans. Internet Technol.}, 22(4):104:1--104:31.

\bibitem[{Kim et~al.(2024)Kim, Park, Jeong, Chan, Xu, McDuff, Lee, Ghassemi, Breazeal, and Park}]{kim_mdagents_2024}
Yubin Kim, Chanwoo Park, Hyewon Jeong, Yik~Siu Chan, Xuhai Xu, Daniel McDuff, Hyeonhoon Lee, Marzyeh Ghassemi, Cynthia Breazeal, and Hae~Won Park. 2024.
\newblock \href {https://openreview.net/forum?id=EKdk4vxKO4} {{MDAgents}: {An} {Adaptive} {Collaboration} of {LLMs} for {Medical} {Decision}-{Making}}.
\newblock In \emph{The Thirty-eighth Annual Conference on Neural Information Processing Systems}.

\bibitem[{Kolt(2025)}]{kolt_governing_2025}
Noam Kolt. 2025.
\newblock \href {https://doi.org/10.48550/arXiv.2501.07913} {Governing {AI} {Agents}}.
\newblock \emph{arXiv preprint}.
\newblock ArXiv:2501.07913 [cs].

\bibitem[{Kotseruba and Tsotsos(2020)}]{kotseruba_cogarch_2020}
Iuliia Kotseruba and John~K. Tsotsos. 2020.
\newblock \href {https://doi.org/10.1007/s10462-018-9646-y} {40 years of cognitive architectures: core cognitive abilities and practical applications}.
\newblock \emph{Artificial Intelligence Review}, 53(1):17--94.

\bibitem[{Krishnan(2025)}]{krishnan_ai_2025}
Naveen Krishnan. 2025.
\newblock \href {https://doi.org/10.48550/arXiv.2503.12687} {{AI} {Agents}: {Evolution}, {Architecture}, and {Real}-{World} {Applications}}.
\newblock \emph{arXiv preprint}.
\newblock ArXiv:2503.12687 [cs].

\bibitem[{Laffont and Martimort(2002)}]{laffont_theory_2002}
Jean-Jacques Laffont and David Martimort. 2002.
\newblock \emph{The {Theory} of {Incentives}: {The} {Principal}-{Agent} {Model}}.
\newblock Princeton University Press, Princeton, N.J.

\bibitem[{Lang et~al.(2024)Lang, Foote, Russell, Dragan, Jenner, and Emmons}]{lang_when_2024}
Leon Lang, Davis Foote, Stuart Russell, Anca Dragan, Erik Jenner, and Scott Emmons. 2024.
\newblock \href {https://openreview.net/forum?id=XcbgkjWSJ7&noteId=7bt8GnwBkH} {When {Your} {AIs} {Deceive} {You}: {Challenges} of {Partial} {Observability} in {Reinforcement} {Learning} from {Human} {Feedback}}.
\newblock In \emph{The Thirty-eighth Annual Conference on Neural Information Processing Systems}.

\bibitem[{Lefouili and Madio(2022)}]{lefouili_economics_2022}
Yassine Lefouili and Leonardo Madio. 2022.
\newblock \href {https://doi.org/10.1007/s10657-022-09728-7} {The economics of platform liability}.
\newblock \emph{European Journal of Law and Economics}, 53(3):319--351.

\bibitem[{Levmore(1990)}]{levmore1990probabilistic}
Saul Levmore. 1990.
\newblock Probabilistic recoveries, restitution, and recurring wrongs.
\newblock \emph{The Journal of Legal Studies}, 19(S2):691--726.

\bibitem[{Li(2025)}]{li_llmagents_2025}
Xinzhe Li. 2025.
\newblock \href {https://aclanthology.org/2025.coling-main.652/} {A {Review} of {Prominent} {Paradigms} for {LLM}-{Based} {Agents}: {Tool} {Use}, {Planning} ({Including} {RAG}), and {Feedback} {Learning}}.
\newblock In \emph{Proceedings of the 31st {International} {Conference} on {Computational} {Linguistics}}, pages 9760--9779, Abu Dhabi, UAE. Association for Computational Linguistics.

\bibitem[{Libert(2025)}]{libert_ai_2025}
Julien Libert. 2025.
\newblock \href {https://www.ceps.eu/an-ai-liability-regulation-would-complete-the-eus-ai-strategy/} {An {AI} {Liability} {Regulation} would complete the {EU}’s {AI} strategy}.
\newblock \emph{CEPS}.

\bibitem[{Lin et~al.(2024)Lin, Ojha, Cai, and Chen}]{lin_strategic_2024}
Ryan~Y. Lin, Siddhartha Ojha, Kevin Cai, and Maxwell Chen. 2024.
\newblock \href {https://openreview.net/forum?id=X9vAImw5Yj} {Strategic {Collusion} of {LLM} {Agents}: {Market} {Division} in {Multi}-{Commodity} {Competitions}}.
\newblock In \emph{Language Gamification - NeurIPS 2024 Workshop}.

\bibitem[{Lior(2024)}]{lior2024holding}
Anat Lior. 2024.
\newblock Holding ai accountable: Addressing the ai-related harms through existing tort doctrines.
\newblock \emph{U. Chi. L. Rev. Online}, page~1.

\bibitem[{Liu et~al.(2025{\natexlab{a}})Liu, Shi, Lyu, Jin, and Faltings}]{liu_copyjudge_2025}
Shunchang Liu, Zhuan Shi, Lingjuan Lyu, Yaochu Jin, and Boi Faltings. 2025{\natexlab{a}}.
\newblock \href {https://doi.org/10.48550/arXiv.2502.15278} {{CopyJudge}: {Automated} {Copyright} {Infringement} {Identification} and {Mitigation} in {Text}-to-{Image} {Diffusion} {Models}}.
\newblock \emph{arXiv preprint}.
\newblock ArXiv:2502.15278 [cs].

\bibitem[{Liu et~al.(2025{\natexlab{b}})Liu, Yao, Jia, Casper, Baracaldo, Hase, Yao, Liu, Xu, Li, Varshney, Bansal, Koyejo, and Liu}]{liu_rethinking_2025}
Sijia Liu, Yuanshun Yao, Jinghan Jia, Stephen Casper, Nathalie Baracaldo, Peter Hase, Yuguang Yao, Chris~Yuhao Liu, Xiaojun Xu, Hang Li, Kush~R. Varshney, Mohit Bansal, Sanmi Koyejo, and Yang Liu. 2025{\natexlab{b}}.
\newblock \href {https://doi.org/10.1038/s42256-025-00985-0} {Rethinking machine unlearning for large language models}.
\newblock \emph{Nature Machine Intelligence}, 7(2):181--194.
\newblock Publisher: Nature Publishing Group.

\bibitem[{Loya et~al.(2023)Loya, Sinha, and Futrell}]{loya_exploring_2023}
Manikanta Loya, Divya Sinha, and Richard Futrell. 2023.
\newblock \href {https://doi.org/10.18653/v1/2023.findings-emnlp.241} {Exploring the {Sensitivity} of {LLMs}' {Decision}-{Making} {Capabilities}: {Insights} from {Prompt} {Variations} and {Hyperparameters}}.
\newblock In \emph{Findings of the {Association} for {Computational} {Linguistics}: {EMNLP} 2023}, pages 3711--3716, Singapore. Association for Computational Linguistics.

\bibitem[{Lu et~al.(2024)Lu, Pan, Chen, Feng, Hu, Peng, and Chen}]{lu_agentlens_2024}
Jiaying Lu, Bo~Pan, Jieyi Chen, Yingchaojie Feng, Jingyuan Hu, Yuchen Peng, and Wei Chen. 2024.
\newblock \href {https://doi.org/10.1109/TVCG.2024.3394053} {{AgentLens}: {Visual} {Analysis} for {Agent} {Behaviors} in {LLM}-based {Autonomous} {Systems}}.
\newblock \emph{IEEE Transactions on Visualization and Computer Graphics}, pages 1--17.
\newblock Conference Name: IEEE Transactions on Visualization and Computer Graphics.

\bibitem[{Lubars and Tan(2019)}]{lubars_ask_2019}
Brian Lubars and Chenhao Tan. 2019.
\newblock \href {https://papers.nips.cc/paper_files/paper/2019/hash/d67d8ab4f4c10bf22aa353e27879133c-Abstract.html} {Ask not what {AI} can do, but what {AI} should do: {Towards} a framework of task delegability}.
\newblock In \emph{Advances in {Neural} {Information} {Processing} {Systems}}, volume~32. Curran Associates, Inc.

\bibitem[{Lyu et~al.(2023)Lyu, Havaldar, Stein, Zhang, Rao, Wong, Apidianaki, and Callison-Burch}]{lyu_faithful_2023}
Qing Lyu, Shreya Havaldar, Adam Stein, Li~Zhang, Delip Rao, Eric Wong, Marianna Apidianaki, and Chris Callison-Burch. 2023.
\newblock \href {https://doi.org/10.18653/v1/2023.ijcnlp-main.20} {Faithful {Chain}-of-{Thought} {Reasoning}}.
\newblock In \emph{Proceedings of the 13th {International} {Joint} {Conference} on {Natural} {Language} {Processing} and the 3rd {Conference} of the {Asia}-{Pacific} {Chapter} of the {Association} for {Computational} {Linguistics} ({Volume} 1: {Long} {Papers})}, pages 305--329, Nusa Dua, Bali. Association for Computational Linguistics.

\bibitem[{Macmillan-Scott and Musolesi(2024)}]{macmillan-scott_irrationality_2024}
Olivia Macmillan-Scott and Mirco Musolesi. 2024.
\newblock \href {https://doi.org/10.1098/rsos.240255} {({Ir})rationality and cognitive biases in large language models}.
\newblock \emph{Royal Society Open Science}, 11(6):240255.
\newblock Publisher: Royal Society.

\bibitem[{Maharana et~al.(2024)Maharana, Lee, Tulyakov, Bansal, Barbieri, and Fang}]{maharana_evaluating_2024}
Adyasha Maharana, Dong-Ho Lee, Sergey Tulyakov, Mohit Bansal, Francesco Barbieri, and Yuwei Fang. 2024.
\newblock \href {https://doi.org/10.18653/v1/2024.acl-long.747} {Evaluating {Very} {Long}-{Term} {Conversational} {Memory} of {LLM} {Agents}}.
\newblock In \emph{Proceedings of the 62nd {Annual} {Meeting} of the {Association} for {Computational} {Linguistics} ({Volume} 1: {Long} {Papers})}, pages 13851--13870, Bangkok, Thailand. Association for Computational Linguistics.

\bibitem[{Mattingly and Cibralic(2025)}]{mattingly_machine_2025}
James Mattingly and Beba Cibralic. 2025.
\newblock \emph{Machine {Agency}}.
\newblock MIT Press.

\bibitem[{Maus et~al.(2023)Maus, Chao, Wong, and Gardner}]{maus_black_2023}
Natalie Maus, Patrick Chao, Eric Wong, and Jacob~R. Gardner. 2023.
\newblock \href {https://openreview.net/forum?id=aI5QPjTRbS} {Black {Box} {Adversarial} {Prompting} for {Foundation} {Models}}.
\newblock In \emph{The Second Workshop on New Frontiers in Adversarial Machine Learning}.

\bibitem[{Mieczkowski et~al.(2025)Mieczkowski, Mon-Williams, Bramley, Lucas, Velez, and Griffiths}]{mieczkowski_tradeoff_2025}
Elizabeth Mieczkowski, Ruaridh Mon-Williams, Neil Bramley, Christopher~G. Lucas, Natalia Velez, and Thomas~L. Griffiths. 2025.
\newblock \href {https://doi.org/10.48550/arXiv.2503.15703} {Predicting {Multi}-{Agent} {Specialization} via {Task} {Parallelizability}}.
\newblock \emph{arXiv preprint}.
\newblock ArXiv:2503.15703 [cs].

\bibitem[{Miehling et~al.(2025)Miehling, Ramamurthy, Varshney, Riemer, Bouneffouf, Richards, Dhurandhar, Daly, Hind, Sattigeri, Wei, Rawat, Gajcin, and Geyer}]{miehling_agentic_2025}
Erik Miehling, Karthikeyan~Natesan Ramamurthy, Kush~R. Varshney, Matthew Riemer, Djallel Bouneffouf, John~T. Richards, Amit Dhurandhar, Elizabeth~M. Daly, Michael Hind, Prasanna Sattigeri, Dennis Wei, Ambrish Rawat, Jasmina Gajcin, and Werner Geyer. 2025.
\newblock \href {https://doi.org/10.48550/arXiv.2503.00237} {Agentic {AI} {Needs} a {Systems} {Theory}}.
\newblock \emph{arXiv preprint}.
\newblock ArXiv:2503.00237 [cs].

\bibitem[{Mirsky et~al.(2022)Mirsky, Carlucho, Rahman, Fosong, Macke, Sridharan, Stone, and Albrecht}]{mirsky_survey_2022}
Reuth Mirsky, Ignacio Carlucho, Arrasy Rahman, Elliot Fosong, William Macke, Mohan Sridharan, Peter Stone, and Stefano~V. Albrecht. 2022.
\newblock \href {https://doi.org/10.1007/978-3-031-20614-6_16} {A {Survey} of {Ad} {Hoc} {Teamwork} {Research}}.
\newblock In \emph{Multi-{Agent} {Systems}}, pages 275--293, Cham. Springer International Publishing.

\bibitem[{Mitchell et~al.(2025)Mitchell, Ghosh, Luccioni, and Pistilli}]{mitchell_fully_2025}
Margaret Mitchell, Avijit Ghosh, Alexandra~Sasha Luccioni, and Giada Pistilli. 2025.
\newblock \href {https://doi.org/10.48550/arXiv.2502.02649} {Fully {Autonomous} {AI} {Agents} {Should} {Not} be {Developed}}.
\newblock \emph{arXiv preprint}.
\newblock ArXiv:2502.02649 [cs].

\bibitem[{Mitchell(2021)}]{mitchell_why_2021}
Neil~J. Mitchell. 2021.
\newblock \emph{Why {Delegate}?}
\newblock Oxford University Press, New York, NY.

\bibitem[{Mitnick(1992)}]{mitnick_theory_1992}
Barry~M. Mitnick. 1992.
\newblock \href {https://papers.ssrn.com/abstract=2164770} {The {Theory} of {Agency} and {Organizational} {Analysis}}.
\newblock In \emph{Ethics and {Agency} {Theory}: {An} {Introduction}}. Oxford University Press, Oxford, United Kingdom.

\bibitem[{Munday(2022)}]{munday_agency_2022}
Roderick Munday. 2022.
\newblock \emph{Agency: {Law} and {Principles}}, 4th edition.
\newblock Oxford University Press, Oxford, United Kingdom ; New York, NY.

\bibitem[{Oldham and Fried(2016)}]{oldham_job_2016}
Greg~R. Oldham and Yitzhak Fried. 2016.
\newblock \href {https://doi.org/10.1016/j.obhdp.2016.05.002} {Job design research and theory: {Past}, present and future}.
\newblock \emph{Organizational Behavior and Human Decision Processes}, 136:20--35.

\bibitem[{Padovan et~al.(2023)Padovan, Martins, and Reed}]{padovan_black_2023}
Paulo~Henrique Padovan, Clarice~Marinho Martins, and Chris Reed. 2023.
\newblock \href {https://doi.org/10.1007/s10506-022-09308-9} {Black is the new orange: how to determine {AI} liability}.
\newblock \emph{Artificial Intelligence and Law}, 31(1):133--167.

\bibitem[{Pan et~al.(2025)Pan, Cemri, Agrawal, Yang, Chopra, Tiwari, Keutzer, Parameswaran, Ramchandran, Klein, Gonzalez, Zaharia, and Stoica}]{pan_whymas_2025}
Melissa~Z. Pan, Mert Cemri, Lakshya~A. Agrawal, Shuyi Yang, Bhavya Chopra, Rishabh Tiwari, Kurt Keutzer, Aditya Parameswaran, Kannan Ramchandran, Dan Klein, Joseph~E. Gonzalez, Matei Zaharia, and Ion Stoica. 2025.
\newblock \href {https://openreview.net/forum?id=wM521FqPvI)} {Why {Do} {Multiagent} {Systems} {Fail}?}
\newblock In \emph{ICLR 2025 Workshop on Building Trust in Language Models and Applications}.

\bibitem[{Park et~al.(2024{\natexlab{a}})Park, Zou, Shaw, Hill, Cai, Morris, Willer, Liang, and Bernstein}]{park_generative_2024}
Joon~Sung Park, Carolyn~Q. Zou, Aaron Shaw, Benjamin~Mako Hill, Carrie Cai, Meredith~Ringel Morris, Robb Willer, Percy Liang, and Michael~S. Bernstein. 2024{\natexlab{a}}.
\newblock \href {https://doi.org/10.48550/arXiv.2411.10109} {Generative {Agent} {Simulations} of 1,000 {People}}.
\newblock \emph{arXiv preprint}.
\newblock ArXiv:2411.10109 [cs].

\bibitem[{Park et~al.(2024{\natexlab{b}})Park, Goldstein, O’Gara, Chen, and Hendrycks}]{park_deception_2024}
Peter~S. Park, Simon Goldstein, Aidan O’Gara, Michael Chen, and Dan Hendrycks. 2024{\natexlab{b}}.
\newblock \href {https://doi.org/10.1016/j.patter.2024.100988} {{AI} deception: {A} survey of examples, risks, and potential solutions}.
\newblock \emph{Patterns}, 5(5).
\newblock Publisher: Elsevier.

\bibitem[{Peigne-Lefebvre et~al.(2025)Peigne-Lefebvre, Kniejski, Sondej, David, Hoelscher-Obermaier, Witt, and Kran}]{peigne-lefebvre_mas_2025}
Pierre Peigne-Lefebvre, Mikolaj Kniejski, Filip Sondej, Matthieu David, Jason Hoelscher-Obermaier, Christian Schroeder~de Witt, and Esben Kran. 2025.
\newblock \href {https://doi.org/10.48550/arXiv.2502.19145} {Multi-{Agent} {Security} {Tax}: {Trading} {Off} {Security} and {Collaboration} {Capabilities} in {Multi}-{Agent} {Systems}}.
\newblock \emph{arXiv preprint}.
\newblock ArXiv:2502.19145 [cs].

\bibitem[{Perez et~al.(2023)Perez, Ringer, Lukosiute, Nguyen, Chen, Heiner, Pettit, Olsson, Kundu, Kadavath, Jones, Chen, Mann, Israel, Seethor, McKinnon, Olah, Yan, Amodei, Amodei, Drain, Li, Tran-Johnson, Khundadze, Kernion, Landis, Kerr, Mueller, Hyun, Landau, Ndousse, Goldberg, Lovitt, Lucas, Sellitto, Zhang, Kingsland, Elhage, Joseph, Mercado, DasSarma, Rausch, Larson, McCandlish, Johnston, Kravec, El~Showk, Lanham, Telleen-Lawton, Brown, Henighan, Hume, Bai, Hatfield-Dodds, Clark, Bowman, Askell, Grosse, Hernandez, Ganguli, Hubinger, Schiefer, and Kaplan}]{perez_discovering_2023}
Ethan Perez, Sam Ringer, Kamile Lukosiute, Karina Nguyen, Edwin Chen, Scott Heiner, Craig Pettit, Catherine Olsson, Sandipan Kundu, Saurav Kadavath, Andy Jones, Anna Chen, Benjamin Mann, Brian Israel, Bryan Seethor, Cameron McKinnon, Christopher Olah, Da~Yan, Daniela Amodei, and 44 others. 2023.
\newblock \href {https://doi.org/10.18653/v1/2023.findings-acl.847} {Discovering {Language} {Model} {Behaviors} with {Model}-{Written} {Evaluations}}.
\newblock In \emph{Findings of the {Association} for {Computational} {Linguistics}: {ACL} 2023}, pages 13387--13434, Stroudsburg, PA, USA. Association for Computational Linguistics.

\bibitem[{Perrier and Bennett(2025)}]{perrier_position_2025}
Elija Perrier and Michael~Timothy Bennett. 2025.
\newblock \href {https://doi.org/10.48550/arXiv.2502.10420} {Position: {Stop} {Acting} {Like} {Language} {Model} {Agents} {Are} {Normal} {Agents}}.
\newblock \emph{arXiv preprint}.
\newblock ArXiv:2502.10420 [cs].

\bibitem[{Piao et~al.(2025)Piao, Yan, Zhang, Li, Yan, Lan, Lu, Zheng, Wang, Zhou, Gao, Xu, Zhang, Rong, Su, and Li}]{piao_agentsociety_2025}
Jinghua Piao, Yuwei Yan, Jun Zhang, Nian Li, Junbo Yan, Xiaochong Lan, Zhihong Lu, Zhiheng Zheng, Jing~Yi Wang, Di~Zhou, Chen Gao, Fengli Xu, Fang Zhang, Ke~Rong, Jun Su, and Yong Li. 2025.
\newblock \href {https://doi.org/10.48550/arXiv.2502.08691} {{AgentSociety}: {Large}-{Scale} {Simulation} of {LLM}-{Driven} {Generative} {Agents} {Advances} {Understanding} of {Human} {Behaviors} and {Society}}.
\newblock \emph{arXiv preprint}.
\newblock ArXiv:2502.08691 [cs].

\bibitem[{Popp(2011)}]{popp_software_2011}
Karl Popp. 2011.
\newblock \href {https://doi.org/10.1109/MS.2011.52} {Software {Industry} {Business} {Models}}.
\newblock \emph{IEEE Software}, 28(4):26--30.
\newblock Conference Name: IEEE Software.

\bibitem[{Price et~al.(2019)Price, Gerke, and Cohen}]{price_potential_2019}
W.~Nicholson Price, Sara Gerke, and I.~Glenn Cohen. 2019.
\newblock \href {https://doi.org/10.1001/jama.2019.15064} {Potential {Liability} for {Physicians} {Using} {Artificial} {Intelligence}}.
\newblock \emph{JAMA}, 322(18):1765.

\bibitem[{Qian et~al.(2025)Qian, Xie, Wang, Liu, Zhu, Xia, Dang, Du, Chen, Yang, Liu, and Sun}]{qian_scaling_2025}
Chen Qian, Zihao Xie, YiFei Wang, Wei Liu, Kunlun Zhu, Hanchen Xia, Yufan Dang, Zhuoyun Du, Weize Chen, Cheng Yang, Zhiyuan Liu, and Maosong Sun. 2025.
\newblock \href {https://openreview.net/forum?id=K3n5jPkrU6} {Scaling {Large} {Language} {Model}-based {Multi}-{Agent} {Collaboration}}.
\newblock In \emph{The Thirteenth International Conference on Learning Representations}.

\bibitem[{Ramakrishnan et~al.(2024)Ramakrishnan, Smith, and Downey}]{ramakrishnan2024us}
Ketan Ramakrishnan, Gregory Smith, and Conor Downey. 2024.
\newblock \href {https://www.rand.org/pubs/research_reports/RRA3084-1.html} {U.{S}. {Tort} {Liability} for {Large}-{Scale} {Artificial} {Intelligence} {Damages}: {A} {Primer} for {Developers} and {Policymakers}}.
\newblock Technical report, RAND Corporation.

\bibitem[{Ratliff et~al.(2019)Ratliff, Dong, Sekar, and Fiez}]{ratliff_perspective_2019}
Lillian~J. Ratliff, Roy Dong, Shreyas Sekar, and Tanner Fiez. 2019.
\newblock \href {https://doi.org/10.1146/annurev-control-053018-023634} {A {Perspective} on {Incentive} {Design}: {Challenges} and {Opportunities}}.
\newblock \emph{Annual Review of Control, Robotics, and Autonomous Systems}, 2(Volume 2, 2019):305--338.
\newblock Publisher: Annual Reviews.

\bibitem[{Reid(1999)}]{reid_liability_1999}
Elspeth Reid. 1999.
\newblock \href {https://doi.org/10.1017/S0020589300063661} {Liability for {Dangerous} {Activities}: {A} {Comparative} {Analysis}}.
\newblock \emph{International \& Comparative Law Quarterly}, 48(4):731--756.

\bibitem[{Rimsky et~al.(2024)Rimsky, Gabrieli, Schulz, Tong, Hubinger, and Turner}]{rimsky_steering_2024}
Nina Rimsky, Nick Gabrieli, Julian Schulz, Meg Tong, Evan Hubinger, and Alexander Turner. 2024.
\newblock \href {https://doi.org/10.18653/v1/2024.acl-long.828} {Steering {Llama} 2 via {Contrastive} {Activation} {Addition}}.
\newblock In \emph{Proceedings of the 62nd {Annual} {Meeting} of the {Association} for {Computational} {Linguistics} ({Volume} 1: {Long} {Papers})}, pages 15504--15522, Bangkok, Thailand. Association for Computational Linguistics.

\bibitem[{Salinas and Morstatter(2024)}]{salinas_butterfly_2024}
Abel Salinas and Fred Morstatter. 2024.
\newblock \href {https://doi.org/10.18653/v1/2024.findings-acl.275} {The {Butterfly} {Effect} of {Altering} {Prompts}: {How} {Small} {Changes} and {Jailbreaks} {Affect} {Large} {Language} {Model} {Performance}}.
\newblock In \emph{Findings of the {Association} for {Computational} {Linguistics}: {ACL} 2024}, pages 4629--4651, Bangkok, Thailand. Association for Computational Linguistics.

\bibitem[{Scheurer et~al.(2024)Scheurer, Balesni, and Hobbhahn}]{scheurer_large_2024}
Jérémy Scheurer, Mikita Balesni, and Marius Hobbhahn. 2024.
\newblock \href {https://openreview.net/forum?id=HduMpot9sJ} {Large {Language} {Models} can {Strategically} {Deceive} their {Users} when {Put} {Under} {Pressure}}.
\newblock In \emph{ICLR 2024 Workshop on Large Language Model (LLM) Agents}.

\bibitem[{Shanahan et~al.(2023)Shanahan, McDonell, and Reynolds}]{Shanahan2023}
Murray Shanahan, Kyle McDonell, and Laria Reynolds. 2023.
\newblock \href {https://doi.org/10.1038/s41586-023-06647-8} {Role play with large language models}.
\newblock \emph{Nature}, 623(7987):493--498.

\bibitem[{Sharkey(2024)}]{sharkey_products_2024}
Catherine~M. Sharkey. 2024.
\newblock \href {https://doi.org/10.52214/stlr.v25i2.12763} {A {Products} {Liability} {Framework} for {AI}}.
\newblock \emph{Columbia Science and Technology Law Review}, 25(2):240--260.
\newblock Number: 2.

\bibitem[{Sharma et~al.(2024)Sharma, Tong, Korbak, Duvenaud, Askell, Bowman, Durmus, Hatfield-Dodds, Johnston, Kravec, Maxwell, McCandlish, Ndousse, Rausch, Schiefer, Yan, Zhang, and Perez}]{sharma_towards_2024}
Mrinank Sharma, Meg Tong, Tomasz Korbak, David Duvenaud, Amanda Askell, Samuel~R. Bowman, Esin Durmus, Zac Hatfield-Dodds, Scott~R. Johnston, Shauna~M. Kravec, Timothy Maxwell, Sam McCandlish, Kamal Ndousse, Oliver Rausch, Nicholas Schiefer, Da~Yan, Miranda Zhang, and Ethan Perez. 2024.
\newblock \href {https://openreview.net/forum?id=tvhaxkMKAn} {Towards {Understanding} {Sycophancy} in {Language} {Models}}.
\newblock In \emph{The Twelfth International Conference on Learning Representations}.

\bibitem[{Shavit et~al.(2023)Shavit, Agarwal, Brundage, O’Keefe, Campbell, Lee, Mishkin, Eloundou, Hickey, Slama, Ahmad, McMillan, Beutel, Passos, and Robinson}]{shavit_practices_2023}
Yonadav Shavit, Sandhini Agarwal, Miles Brundage, Steven Adler~Cullen O’Keefe, Rosie Campbell, Teddy Lee, Pamela Mishkin, Tyna Eloundou, Alan Hickey, Katarina Slama, Lama Ahmad, Paul McMillan, Alex Beutel, Alexandre Passos, and David~G. Robinson. 2023.
\newblock \href {https://www.semanticscholar.org/paper/Practices-for-Governing-Agentic-AI-Systems-Shavit-Agarwal/0002c42e8d7bfeafc431c4ed9f6318f223bbf58b} {Practices for {Governing} {Agentic} {AI} {Systems}}.

\bibitem[{Soder et~al.(2024)Soder, Smakman, Dunlop, Pan, Swaroop, and Kolt}]{soder_levels_2024}
Lisa Soder, Julia Smakman, Connor Dunlop, Weiwei Pan, Siddharth Swaroop, and Noam Kolt. 2024.
\newblock \href {https://openreview.net/forum?id=EH6SmoChx9} {Levels of {Autonomy}: {Liability} in the age of {AI} {Agents}}.
\newblock In \emph{Workshop on Socially Responsible Language Modelling Research}.

\bibitem[{Spence(1973)}]{spence1973job}
Michael Spence. 1973.
\newblock \href {https://doi.org/10.2307/1882010} {Job market signaling}.
\newblock \emph{The Quarterly Journal of Economics}, 87:355--374.

\bibitem[{Sterz et~al.(2024)Sterz, Baum, Biewer, Hermanns, Lauber-Rönsberg, Meinel, and Langer}]{sterz_quest_2024}
Sarah Sterz, Kevin Baum, Sebastian Biewer, Holger Hermanns, Anne Lauber-Rönsberg, Philip Meinel, and Markus Langer. 2024.
\newblock \href {https://doi.org/10.1145/3630106.3659051} {On the {Quest} for {Effectiveness} in {Human} {Oversight}: {Interdisciplinary} {Perspectives}}.
\newblock In \emph{Proceedings of the 2024 {ACM} {Conference} on {Fairness}, {Accountability}, and {Transparency}}, {FAccT} '24, pages 2495--2507, New York, NY, USA. Association for Computing Machinery.

\bibitem[{Street(2024)}]{street_llm_2024}
Winnie Street. 2024.
\newblock \href {https://doi.org/10.48550/arXiv.2405.08154} {{LLM} {Theory} of {Mind} and {Alignment}: {Opportunities} and {Risks}}.
\newblock \emph{arXiv preprint}.
\newblock ArXiv:2405.08154 [cs].

\bibitem[{Sumers et~al.(2023)Sumers, Yao, Narasimhan, and Griffiths}]{sumers_cognitive_2023}
Theodore Sumers, Shunyu Yao, Karthik Narasimhan, and Thomas Griffiths. 2023.
\newblock \href {https://openreview.net/forum?id=1i6ZCvflQJ} {Cognitive {Architectures} for {Language} {Agents}}.
\newblock \emph{Transactions on Machine Learning Research}.

\bibitem[{Sykes(1984)}]{sykes_economics_1984}
Alan Sykes. 1984.
\newblock \href {https://openyls.law.yale.edu/handle/20.500.13051/16285} {The {Economics} of {Vicarious} {Liability}}.
\newblock \emph{Yale Law Journal}.
\newblock Accepted: 2021-11-26T12:26:37Z.

\bibitem[{Tang et~al.(2024)Tang, Zou, Zhang, Li, Zhao, Zhang, Cohan, and Gerstein}]{tang_medagents_2024}
Xiangru Tang, Anni Zou, Zhuosheng Zhang, Ziming Li, Yilun Zhao, Xingyao Zhang, Arman Cohan, and Mark Gerstein. 2024.
\newblock \href {https://doi.org/10.18653/v1/2024.findings-acl.33} {{MedAgents}: {Large} {Language} {Models} as {Collaborators} for {Zero}-shot {Medical} {Reasoning}}.
\newblock In \emph{Findings of the {Association} for {Computational} {Linguistics}: {ACL} 2024}, pages 599--621, Bangkok, Thailand. Association for Computational Linguistics.

\bibitem[{Thomas et~al.(2017)Thomas, Crook, and Edelman}]{thomas_credit_2017}
Lyn Thomas, Jonathan Crook, and David Edelman. 2017.
\newblock \href {https://doi.org/10.1137/1.9781611974560} {\emph{Credit {Scoring} and {Its} {Applications}}}, second edition.
\newblock Mathematics in {Industry}. Society for Industrial and Applied Mathematics.

\bibitem[{Tran et~al.(2025)Tran, Dao, Nguyen, Pham, O'Sullivan, and Nguyen}]{tran_multiagent_2025}
Khanh-Tung Tran, Dung Dao, Minh-Duong Nguyen, Quoc-Viet Pham, Barry O'Sullivan, and Hoang~D. Nguyen. 2025.
\newblock \href {https://doi.org/10.48550/arXiv.2501.06322} {Multi-{Agent} {Collaboration} {Mechanisms}: {A} {Survey} of {LLMs}}.
\newblock \emph{arXiv preprint}.
\newblock ArXiv:2501.06322 [cs].

\bibitem[{Trivedi et~al.(2024)Trivedi, Chandak, Muresanu, Zhu, Sarkar, Leibo, Hadfield-Menell, and Hadfield}]{trivedi_altared_2024}
Rakshit Trivedi, Nikhil Chandak, Andrei~Ioan Muresanu, Shuhui Zhu, Atrisha Sarkar, Joel~Z. Leibo, Dylan Hadfield-Menell, and Gillian~K. Hadfield. 2024.
\newblock \href {https://openreview.net/forum?id=Gd6QrBLHBN} {Altared {Environments}: {The} {Role} of {Normative} {Infrastructure} in {AI} {Alignment}}.
\newblock In \emph{Agentic Markets Workshop at ICML 2024}.

\bibitem[{Tuan et~al.(2021)Tuan, Pryor, Chen, Getoor, and Wang}]{tuan_local_2021}
Yi-Lin Tuan, Connor Pryor, Wenhu Chen, Lise Getoor, and William~Yang Wang. 2021.
\newblock \href {https://proceedings.neurips.cc/paper/2021/hash/03b92cd507ff5870df0db7f074728830-Abstract.html} {Local {Explanation} of {Dialogue} {Response} {Generation}}.
\newblock In \emph{Advances in {Neural} {Information} {Processing} {Systems}}, volume~34, pages 404--416. Curran Associates, Inc.

\bibitem[{Turner(2018)}]{turner2018robot}
Jacob Turner. 2018.
\newblock \emph{Robot {Rules}: {Regulating} {Artificial} {Intelligence}}.
\newblock Palgrave Macmillan, New York, NY.

\bibitem[{Vardi and Weitz(2016)}]{vardi_misbehavior_2016}
Yoav Vardi and Ely Weitz. 2016.
\newblock \emph{Misbehavior in {Organizations}: {A} {Dynamic} {Approach}}, 2nd edition.
\newblock Routledge.

\bibitem[{Wang et~al.(2024{\natexlab{a}})Wang, Rahman, Durugkar, Liebman, and Stone}]{wang_nagent_2024}
Caroline Wang, Arrasy Rahman, Ishan Durugkar, Elad Liebman, and Peter Stone. 2024{\natexlab{a}}.
\newblock \href {https://proceedings.neurips.cc/paper_files/paper/2024/hash/cabf611498431ad89a85ace75f790d93-Abstract-Conference.html} {N-agent {Ad} {Hoc} {Teamwork}}.
\newblock \emph{Advances in Neural Information Processing Systems}, 37:111832--111862.

\bibitem[{Wang et~al.(2025)Wang, Wang, Tang, Li, Chen, Liang, and He}]{wang_all_2025}
Qian Wang, Tianyu Wang, Zhenheng Tang, Qinbin Li, Nuo Chen, Jingsheng Liang, and Bingsheng He. 2025.
\newblock \href {https://openreview.net/forum?id=mSKj2bCWRr} {All {It} {Takes} {Is} {One} {Prompt}: {An} {Autonomous} {LLM}-{MA} {System}}.
\newblock In \emph{ICLR 2025 Workshop on Foundation Models in the Wild}.

\bibitem[{Wang et~al.(2024{\natexlab{b}})Wang, Chen, Yuan, Zhang, Li, Peng, and Ji}]{wang_executable_2024}
Xingyao Wang, Yangyi Chen, Lifan Yuan, Yizhe Zhang, Yunzhu Li, Hao Peng, and Heng Ji. 2024{\natexlab{b}}.
\newblock Executable code actions elicit better {LLM} agents.
\newblock In \emph{Proceedings of the 41st {International} {Conference} on {Machine} {Learning}}, volume 235 of \emph{{ICML}'24}, pages 50208--50232, Vienna, Austria. JMLR.org.

\bibitem[{Wang et~al.(2024{\natexlab{c}})Wang, Zhong, Huang, Shi, Yang, Chen, Li, Ma, Wang, and Zheng}]{wang_agents_2024}
Yanlin Wang, Wanjun Zhong, Yanxian Huang, Ensheng Shi, Min Yang, Jiachi Chen, Hui Li, Yuchi Ma, Qianxiang Wang, and Zibin Zheng. 2024{\natexlab{c}}.
\newblock \href {https://doi.org/10.48550/arXiv.2409.09030} {Agents in {Software} {Engineering}: {Survey}, {Landscape}, and {Vision}}.
\newblock \emph{arXiv preprint}.
\newblock ArXiv:2409.09030 [cs].

\bibitem[{Wei~Jie et~al.(2024)Wei~Jie, Satapathy, Goh, and Cambria}]{wei_jie_how_2024}
Yeo Wei~Jie, Ranjan Satapathy, Rick Goh, and Erik Cambria. 2024.
\newblock \href {https://doi.org/10.18653/v1/2024.findings-naacl.138} {How {Interpretable} are {Reasoning} {Explanations} from {Prompting} {Large} {Language} {Models}?}
\newblock In \emph{Findings of the {Association} for {Computational} {Linguistics}: {NAACL} 2024}, pages 2148--2164, Mexico City, Mexico. Association for Computational Linguistics.

\bibitem[{Weidinger et~al.(2023)Weidinger, Rauh, Marchal, Manzini, Hendricks, Mateos-Garcia, Bergman, Kay, Griffin, Bariach, Gabriel, Rieser, and Isaac}]{weidinger_sociotechnical_2023}
Laura Weidinger, Maribeth Rauh, Nahema Marchal, Arianna Manzini, Lisa~Anne Hendricks, Juan Mateos-Garcia, Stevie Bergman, Jackie Kay, Conor Griffin, Ben Bariach, Iason Gabriel, Verena Rieser, and William Isaac. 2023.
\newblock \href {https://doi.org/10.48550/arXiv.2310.11986} {Sociotechnical {Safety} {Evaluation} of {Generative} {AI} {Systems}}.
\newblock \emph{arXiv preprint}.
\newblock ArXiv:2310.11986 [cs].

\bibitem[{Williams et~al.(2025)Williams, Carroll, Narang, Weisser, Murphy, and Dragan}]{williams_targeted_2025}
Marcus Williams, Micah Carroll, Adhyyan Narang, Constantin Weisser, Brendan Murphy, and Anca Dragan. 2025.
\newblock \href {https://openreview.net/forum?id=Wf2ndb8nhf} {On {Targeted} {Manipulation} and {Deception} when {Optimizing} {LLMs} for {User} {Feedback}}.
\newblock In \emph{The Thirteenth International Conference on Learning Representations}.

\bibitem[{Xiang et~al.(2024)Xiang, Zheng, Li, Hong, Li, Xie, Zhang, Xiong, Xie, Yang, Song, and Li}]{xiang_guardagent_2024}
Zhen Xiang, Linzhi Zheng, Yanjie Li, Junyuan Hong, Qinbin Li, Han Xie, Jiawei Zhang, Zidi Xiong, Chulin Xie, Carl Yang, Dawn Song, and Bo~Li. 2024.
\newblock \href {https://doi.org/10.48550/arXiv.2406.09187} {{GuardAgent}: {Safeguard} {LLM} {Agents} by a {Guard} {Agent} via {Knowledge}-{Enabled} {Reasoning}}.
\newblock \emph{arXiv preprint}.
\newblock ArXiv:2406.09187 [cs].

\bibitem[{Xiao et~al.(2025)Xiao, Sun, Luo, and Wang}]{xiao_tradingagents_2025}
Yijia Xiao, Edward Sun, Di~Luo, and Wei Wang. 2025.
\newblock \href {https://openreview.net/forum?id=4QPrXwMQt1} {{TradingAgents}: {Multi}-{Agents} {LLM} {Financial} {Trading} {Framework}}.
\newblock In \emph{The First MARW: Multi-Agent AI in the Real World Workshop at AAAI 2025}.

\bibitem[{Xie et~al.(2024)Xie, Chen, Jia, Ye, Lai, Shu, Gu, Bibi, Hu, Jurgens, Evans, Torr, Ghanem, and Li}]{xie_can_2024}
Chengxing Xie, Canyu Chen, Feiran Jia, Ziyu Ye, Shiyang Lai, Kai Shu, Jindong Gu, Adel Bibi, Ziniu Hu, David Jurgens, James Evans, Philip Torr, Bernard Ghanem, and Guohao Li. 2024.
\newblock \href {https://openreview.net/forum?id=CeOwahuQic&referrer=%5Bthe%20profile%20of%20Ziniu%20Hu%5D(%2Fprofile%3Fid%3D~Ziniu_Hu1)} {Can {Large} {Language} {Model} {Agents} {Simulate} {Human} {Trust} {Behavior}?}
\newblock In \emph{The Thirty-eighth Annual Conference on Neural Information Processing Systems}.

\bibitem[{Xu et~al.(2024)Xu, Qi, Guo, Wang, Wang, Zhang, and Xu}]{xu_knowledge_2024}
Rongwu Xu, Zehan Qi, Zhijiang Guo, Cunxiang Wang, Hongru Wang, Yue Zhang, and Wei Xu. 2024.
\newblock \href {https://doi.org/10.18653/v1/2024.emnlp-main.486} {Knowledge {Conflicts} for {LLMs}: {A} {Survey}}.
\newblock In \emph{Proceedings of the 2024 {Conference} on {Empirical} {Methods} in {Natural} {Language} {Processing}}, pages 8541--8565, Miami, Florida, USA. Association for Computational Linguistics.

\bibitem[{Zambonelli et~al.(2003)Zambonelli, Jennings, and Wooldridge}]{zambonelli_developing_2003}
Franco Zambonelli, Nicholas~R. Jennings, and Michael Wooldridge. 2003.
\newblock \href {https://doi.org/10.1145/958961.958963} {Developing multiagent systems: {The} {Gaia} methodology}.
\newblock \emph{ACM Trans. Softw. Eng. Methodol.}, 12(3):317--370.

\bibitem[{Zhang et~al.(2024)Zhang, Cai, Zuo, Luan, Wang, Hou, Zhang, Wei, Sun, Sun, Sun, and Dong}]{zhang_fusion_2024}
Yedi Zhang, Yufan Cai, Xinyue Zuo, Xiaokun Luan, Kailong Wang, Zhe Hou, Yifan Zhang, Zhiyuan Wei, Meng Sun, Jun Sun, Jing Sun, and Jin~Song Dong. 2024.
\newblock \href {https://doi.org/10.48550/arXiv.2412.06512} {The {Fusion} of {Large} {Language} {Models} and {Formal} {Methods} for {Trustworthy} {AI} {Agents}: {A} {Roadmap}}.
\newblock \emph{arXiv preprint}.
\newblock ArXiv:2412.06512 [cs].

\bibitem[{Zheng et~al.(2023)Zheng, Xu, Choudhry, Chen, Li, and Huang}]{zheng_synergizing_2023}
Qingxiao Zheng, Zhongwei Xu, Abhinav Choudhry, Yuting Chen, Yongming Li, and Yun Huang. 2023.
\newblock \href {https://doi.org/10.48550/arXiv.2310.15065} {Synergizing {Human}-{AI} {Agency}: {A} {Guide} of 23 {Heuristics} for {Service} {Co}-{Creation} with {LLM}-{Based} {Agents}}.
\newblock \emph{arXiv preprint}.
\newblock ArXiv:2310.15065 [cs].

\bibitem[{Zhuo et~al.(2024)Zhuo, Zhang, Fang, Duan, Lin, and Chen}]{zhuo_prosa_2024}
Jingming Zhuo, Songyang Zhang, Xinyu Fang, Haodong Duan, Dahua Lin, and Kai Chen. 2024.
\newblock \href {https://doi.org/10.18653/v1/2024.findings-emnlp.108} {{ProSA}: {Assessing} and {Understanding} the {Prompt} {Sensitivity} of {LLMs}}.
\newblock In \emph{Findings of the {Association} for {Computational} {Linguistics}: {EMNLP} 2024}, pages 1950--1976, Miami, Florida, USA. Association for Computational Linguistics.

\bibitem[{Čerka et~al.(2015)Čerka, Grigienė, and Sirbikytė}]{vcerka2015liability}
Paulius Čerka, Jurgita Grigienė, and Gintarė Sirbikytė. 2015.
\newblock \href {https://doi.org/10.1016/j.clsr.2015.03.008} {Liability for damages caused by artificial intelligence}.
\newblock \emph{Computer Law \& Security Review}, 31(3):376--389.

\end{thebibliography}

\end{document}